\documentclass[twocolumn,10pt]{IEEEtran}
\usepackage{amssymb}
\usepackage{theorem,amsbsy,pifont,verbatim,amsfonts,amssymb,color}
\usepackage{xspace,epsfig,graphicx,subfigure,latexsym,fancyhdr}
\usepackage{multirow}
\usepackage{epstopdf}
\usepackage{makecell}
\usepackage{bm}
\usepackage{booktabs}
\usepackage{indentfirst}
\usepackage{amsmath}
\usepackage{diagbox}
\usepackage{enumerate}
\usepackage{float}
\usepackage{url}
\usepackage{mdwmath}
\usepackage{mdwtab}
\usepackage{amsmath}
\usepackage{xcolor}
\usepackage{caption}
\usepackage{graphfig}
\usepackage{algorithm}
\usepackage{algpseudocode}
\usepackage[colorlinks,citecolor=blue,linkcolor=blue,urlcolor=blue]{hyperref}

\def\0{\boldsymbol 0}
\def\1{\boldsymbol 1}

\newtheorem{remark}{Remark}

\allowdisplaybreaks

\ifCLASSINFOpdf

\else

\fi

\usepackage{stfloats}

\begin{document}
	
	\title{FAS-ARIS: Turning Multipath Challenges Into Localization Opportunities}
	
	\author{{
			Hua Chen, 
			Tao Gong, 
			Tuo Wu, 
			Maged Elkashlan, 
			Baiyang Liu,\\ 
			Chan-Byoung Chae, \emph{Fellow, IEEE}, 
			Kin-Fai Tong, \emph{Fellow, IEEE}, and 
			Kai-Kit Wong, \emph{Fellow}, \emph{IEEE}
			\vspace{-5mm}
		}
		
		\thanks{  (\emph{Corresponding author: Tuo Wu and Kai-Kit Wong}.)}
		\thanks{This work was supported by the Zhejiang Provincial Natural Science Foundation of China under Grant LY23F010003, and by the ``Pioneer" and ``Leading Goose" R \& D Program of Zhejiang Province under Grant 2024C01105, and by the China Scholarship Council under Grant 202408330215. Hua Chen and  Tao Gong are Faculty of Electrical Engineering and Computer Science, Ningbo University, Ningbo 315211, China. (E-mail: $\rm 2311100045@nbu.edu.cn; dkchenhua0714@hotmail.com$). T. Wu is with the School of Electrical and Electronic Engineering, Nanyang Technological University 639798, Singapore (E-mail: $\rm  tuo.wu@qmul.ac.uk$). M. Elkashlan is with the School of Electronic Engineering and Computer Science at Queen Mary University of London, London E1 4NS, U.K. (E-mail: $\rm maged.elkashlan@qmul.ac.uk$).  B. Liu and  K. F. Tong are with the School of Science and Technology, Hong Kong Metropolitan University, Hong Kong SAR, China. (E-mail: $\rm \{byliu,ktong\}@hkmu.edu.hk$). C.-B. Chae is with the School of Integrated Technology, Yonsei University, Seoul 03722 Korea. (E-mail: $\rm cbchae@yonsei.ac.kr$).  K.-K. Wong is with the Department of Electronic and Electrical Engineering, University College London, WC1E 6BT London, U.K., and also with the Yonsei Frontier Laboratory, Yonsei University, Seoul 03722, South Korea (E-mail: $\rm kai\text{-}kit.wong@ucl.ac.uk$).} 
	}
	
	\markboth{}
	{Shell \MakeLowercase{\textit{et al.}}: Bare Demo of IEEEtran.cls for Journals}
	
	\maketitle
	
	\begin{abstract}
		Traditional single-input single-output (SISO) systems face fundamental limitations in achieving accurate three-dimensional (3D) localization due to limited spatial degrees of freedom (DoF) and the adverse impact of multipath propagation. This paper proposes a novel fluid antenna system (FAS)-active reconfigurable intelligent surface (ARIS) framework that transforms multipath effects from a hindrance into a resource for enhanced localization. By synergistically combining the signal amplification capabilities of ARIS with the spatial diversity enabled by FAS, the proposed system achieves robust 3D user equipment (UE) positioning---without relying on auxiliary information such as time-of-arrival (ToA) or frequency diversity. The system exploits both line-of-sight (LoS) and non-line-of-sight (NLoS) components through a tailored signal decoupling strategy. We design novel UE pilot sequences and ARIS phase configurations to effectively separate LoS and NLoS channels, enabling independent parameter estimation. A multi-stage estimation algorithm is then applied: the multiple signal classification (MUSIC) algorithm estimates angle-of-arrival (AoA) from the direct path, while maximum likelihood estimation with interior-point refinement recovers cascaded channel parameters from the reflected path. Finally, geometric triangulation using least-squares estimation determines the UE's 3D position based on the extracted AoA information. Comprehensive performance analysis, including the derivation of Cram\'{e}r-Rao bounds for both channel and position estimation, establishes theoretical benchmarks. Simulation results confirm that the proposed FAS-ARIS framework achieves near-optimal localization accuracy while maintaining robustness in rich multipath environments---effectively turning conventional localization challenges into advantages.
	\end{abstract}
	
	
	\IEEEpeerreviewmaketitle
	
	\vspace{-2mm}
	\section{Introduction}\label{sec:1}
	\subsection{Backgrounds}
	\IEEEPARstart{A}{s the} development of sixth-generation (6G) networks accelerates, the limitations of traditional antenna systems from the fifth generation (5G), such as massive multiple input multiple output (MIMO) technology, are becoming evident, particularly in their inability to satisfy the stringent quality of service (QoS) requirements anticipated in 6G \cite{Tariq-2020,10054381}. In response, fluid antenna systems (FAS) are gaining prominence as a crucial innovation for next-generation wireless communications. These systems dynamically adjust antenna positions, greatly enhancing transmission rates and optimizing spatial resource utilization for 6G networks \cite{wong2022bruce}.
	
	Moving beyond traditional fixed-position antennas (FPAs), FAS introduces a revolutionary approach to wireless communications \cite{9131873}. {The work in \cite{wong2021FAS} innovatively proposed the concept of FAS and provided a comprehensive description of their hardware structure and working principles.} Unlike FPA-based systems, FAS utilizes highly reconfigurable radiating structures on a shared radio frequency (RF) chain to achieve radiating position change with or without physical movement. FAS can be implemented using movable arrays \cite{basbug2017design}, liquid-based antennas \cite{Shamim-2025,shen2024design}, metamaterial-based structures \cite{Liu-2025arxiv}, reconfigurable pixels \cite{zhang2024pixel} etc. FAS can effectively mimick the performance of large-scale arrays (but with spatial correlation), significantly reducing hardware costs through adaptive reconfiguration while optimizing system performance. This introduces new spatial degrees of freedom (DoFs) to dynamically adapt to the channel condition for enhancing system performance. 
	
	Since its emergence, numerous efforts have been spent on understanding the fundamental performance limits of FAS channels \cite{Khammassi2023,New2023fluid,Alvim2023on}. Many studies have further considered the use of FAS for a wide range of applications and illustrated promising results. For example, the works in \cite{5} and \cite{6} applied FAS for improving wireless powered communication systems. FAS has also found its use for proactive monitoring systems in \cite{7}. Additionally, the authors of \cite{11} employed FAS to assist non-orthogonal multiple access (NOMA) short packet communication systems. Spectrum sensing in cognitive radio networks is another example that can benefit greatly from FAS \cite{12}. Most recently, FAS has been investigated to improve the performance of integrated sensing and communication (ISAC) \cite{13,TNSE5}. Channel estimation methods for FAS have also been developed in \cite{Skouroumounis2023fluid}.
	
	Meanwhile reconfigurable intelligent surfaces {(RISs)} have recently emerged as an enabling technology for 6G \cite{Huang-2019,Renzo-2020}, for their capability of engineering the radio environment by intelligent reflection. RIS has the ability to improve coverage, localization, and sensing. By precisely controlling both the amplitude and phase of the reflected signals, RIS can effectively reshape the wireless propagation environment. This fine-grained manipulation enables accurate reconstruction of the wireless channel, thereby enhancing signal quality, improving link reliability, and supporting advanced communication functions such as beamforming, localization, and interference suppression. RIS has achieved breakthrough applications in various fields. {The work in \cite{TNSE1} proposed a RIS-assisted secure air-to-ground communication framework, where the RIS enhanced the confidentiality of UAV-to-ground transmissions. The work in \cite{TNSE2} proposed a hybrid discrete-continuous action algorithm, enabling user tasks to be executed locally or offloaded via RIS to cooperative devices or an edge server. The work in \cite{TNSE3} investigated the coordination of sensing and computation offloading in a RIS-assisted, BS-centric symbiotic radio (SR) system. A priority-aware, user-traffic-dependent, group-based multi-hop routing scheme for RIS-assisted millimeter-wave (mmWave) device-to-device (D2D) communication networks with spatially correlated channels was proposed in \cite{TNSE4}}. But traditional RIS is passive and many existing studies \cite{21,22,23} focus on repairing the non-line-of-sight (NLoS) channels to enhance performance. In this case, however, the achievable performance is rather limited due to the `multiplicative fading' or `double fading' effect \cite{24}. 
	
	To overcome this, a new RIS architecture known as active RIS (ARIS) has been proposed \cite{25}, which can actively reflect and amplify incoming signals by integrating reflective amplifiers into their unit cells. {The work in \cite{26} provided a comprehensive introduction to the structure and working methods of ARIS}. Unlike passive RIS that merely adjust the phase of reflected signals, ARIS is equipped with active reflective amplifiers that also amplify the signals. With additional power consumption, ARIS can effectively compensate for severe path loss of the {reflected} link \cite{27}, thereby achieving a genuine enhancement in channel capacity \cite{28}.
	
\begin{table*}[ht]
	\centering
	\caption{FAS and RIS contributions/advantages to location systems.}\label{tab:comprehensive_fas}
	\begin{small}

			\begin{tabular}{>{\raggedright\arraybackslash}p{2.7cm}|>{\raggedright\arraybackslash}p{2.8cm}|>{\raggedright\arraybackslash}p{2.8cm}|>{\raggedright\arraybackslash}p{6.5cm}}
				\hline
				\textbf{FAS/RIS} & \textbf{References} & \textbf{Purpose and application system} & \textbf{Key contributions/advantages}\\
				\hline\hline
				
				\textbf{FAS} 
				& \cite{basbug2017design, zhang2024pixel, Liu-2025arxiv} 
				& Antenna architecture and design 
				& Reconfigurable architecture, fast tuning, low sidelobes, deep nulls, diverse beam patterns, hardware validation, scalable pixel design, and multi-user support without precoding.\\
				\cline{2-4}
				& \cite{Khammassi2023, Espinosa-2024} 
				& Channel modeling for FAS 
				& Accurate, tractable channel models with spatial/block correlation, enabling realistic performance prediction and offering insights into design limits.\\
				\cline{2-4}
				& \cite{New2023fluid, Vega2023asimple, Alvim2023on} 
				& Performance analysis and asymptotic methods 
				& Compact diversity gain, analytical comparison with MRC, nonlinearity impact analysis, asymptotic matching, and performance under Nakagami-$m$ fading.\\
				\cline{2-4}
				& \cite{new2023information} 
				& Joint optimization for MIMO-FAS 
				& 2D MIMO-FAS design with joint port selection and beamforming, improving capacity, diversity-multiplexing tradeoff, and high-SNR reliability.\\
				\hline
				
				\textbf{RIS} 
				& \cite{31, 32, 33} 
				& 3D positioning in SISO systems 
				& High-accuracy, robust localization via scalable, low-complexity algorithms; effective in near/far-field, wideband scenarios using advanced modeling.\\
				\cline{2-4} 
				\hline
				
				\textbf{ARIS} 
				& \cite{26, 27, 28} 
				& Expanding passive RIS capabilities 
				& Active RIS amplifies signals to enhance link quality, improve spectral/energy efficiency, reduce element count, support compact deployment, and strengthen physical layer security.\\
				\hline
			\end{tabular}
		
	\end{small}
\end{table*}
	\vspace{-3mm}
	\subsection{Related Works}
	{
	 In recent years, FASs have been extensively investigated and are increasingly recognized as a promising enabler for tackling diverse challenges in wireless communications, with numerous studies exploring their potential in both theoretical and practical domains. In the terms of direction-of-arrival (DOA) estimation, a sparse FAS was proposed in \cite{DOA1} to enhance DOA estimation performance in mmWave environments. Further advancing this concept, the work in \cite{DOA3} introduced a novel DOA estimation framework based on FAS operating under constrained movement time, which was tailored formunication systems and aims to meet the demanding requirements of high-precision and low-latency sensing in dynamic and complex scenarios. In the field of localization, the work in \cite{wu2024fluid} characterized FAS as a reconfigurable paradigm offering flexibility in both shape and positional adaptation, while providing a comprehensive survey of implementation techniques and applications spanning SWIPT, ISAC, NOMA, RIS, PLS, MEC and others. Meanwhile, a novel scalable FAS was introduced in \cite{DOA2}, which dynamically adjusts its aperture to achieve high-accuracy source localization without requiring prior knowledge of near-field or far-field conditions. In the field of MIMO system, the work in \cite{Wang-2024ai} conceptualized FAS as a MIMO extension that incorporates antenna position as an additional DOF, while proposing an AI-driven framework for joint CSI acquisition and position precoding optimization. Besides, the work in \cite{new2023information} analyzed MIMO-FAS with 2D FAS, introducd $q$-outage capacity as a performance metric, and derived the multiplexing trade-off. In the field of channel estimation, a low-sample sparse channel reconstruction method for FAS in mmWave systems was proposed in \cite{xu2024channel}, enabling near-perfect CSI acquisition with limited antenna switching. The work in \cite{zhang2023successive} developed a successive Bayesian reconstructor that treats channels as stochastic processes, achieving high accuracy under both mismatched and matched channel models. While the work in \cite{Xu-2025ce} formulated channel estimation as a sparse Bayesian learning problem and introduced an improved variant for enhanced reconstruction accuracy. The work in \cite{10751774} analyzed FAS channel estimation under an electromagnetic-compliant model, derived the trade-off between reconstruction accuracy and sampling density, and proposed suboptimal sampling strategies with MLE error bounds. Additionally, the authors of \cite{New2024aTutorial} provided a comprehensive review of FAS as a position and shape-flexible technology, emphasizing physics-consistent modeling, machine learning-based channel estimation, and channel hardening. In the terms of ISAC, the work in \cite{14} investigated the fluid antenna-assisted ISAC, optimizing beamforming and antenna positions with alternating optimization and showing performance improvement under sensing constraints. Furthermore, a number of works have been devoted to structural innovations in FAS, encompassing in-depth analyses as well as the development of novel architectures. The authors of \cite{Lu-2025} established a unified framework grounded in eigenmode theory and modal parity, conceptualizing FAS as antennas featuring controllable boundary conditions, variable spatial occupation, and reconfigurable feeding mechanisms. FAS serves as the hardware foundation by providing position-flexible antenna elements, where fluid antenna multiple access (FAMA) leverages this flexibility to suppress interference and support massive multiuser connectivity. Specifically, the work in \cite{wong2022FAMA} introduced FAMA with a scheme that exploits deep interference fades to enhance outage performance and system capacity. Further advancing this line of work, the slow-FAMA was investigated and closed-form outage expressions as well as quantifying multiplexing gains under practical switching constraints were derived in \cite{wong2023sFAMA}. To facilitate tractable analysis of highly correlated FAS channels, a block-correlation model was proposed in \cite{Espinosa-2024}. The work in \cite{Vega2023asimple} analyzed single-antenna FAS operating over Nakagami-m fading channels using asymptotic matching, demonstrating performance comparable or superior to maximal ratio combining (MRC).}

	By combining FAS and ARIS, FAS-ARIS systems could see notable improvements in signal strength and communication performance. The use of ARIS will be effective in mitigating the multiplicative path loss, especially in situations in which direct signal paths are obstructed, while FAS ensures to exploit fluctuation of the fading channel for diversity in space. In fact, the synergy of FAS and generally RIS-assisted systems has recently been investigated \cite{10539238,3,Yao2025RIS}. The authors of \cite{Zhu-2025ris} further studied index modulation designs using FAS in RIS-assisted communication systems. Comparatively, a few results are available for FAS-ARIS systems. In \cite{29}, the authors tried to compare between FAS and ARIS while \cite{30} investigated the outage probability for FAS-ARIS systems. 
	
	Despite the above-mentioned results, FAS-ARIS systems are still not well known and little is understood to their potential for channel estimation and localization, which motivates this work. Current single-input single-output (SISO) systems face fundamental limitations in achieving positioning without auxiliary information. RIS has been proposed to aid localization. For example, {the work in \cite{31} employed RIS to assist localization in wideband SISO systems, while \cite{32} examined the impact of Doppler on the spatial-wideband effect. Also, The work in \cite{33} focused on SISO localization in multipath near-field environments. All of these methods require certain forms of auxiliary information (such as delay) to achieve accurate localization.}
	
	The integration of FAS and ARIS presents unique opportunities for localization. In FPA-ARIS systems, the limited spatial parameters make three-dimensional (3D) localization challenging due to insufficient DoF. The introduction of FAS fundamentally addresses this limitation by providing additional spatial DoF through dynamic antenna positioning. Also, ARIS amplifies the reflected signals, significantly enhancing the quality of the cascaded channel that would otherwise suffer from severe multiplicative fading. This amplification is particularly beneficial for localization, as it improves the signal-to-noise ratio (SNR) of the reflected path, enabling more accurate angle-of-arrival (AoA) estimation. The synergy between FAS and ARIS therefore creates a powerful framework where FAS provides the necessary spatial diversity while ARIS ensures sufficient signal strength for reliable parameter estimation.
	
	\subsection{Our Contributions}
	But FAS-ARIS localization introduces significant technical challenges. The cascaded nature of ARIS channels inherently complicates parameter estimation, as the desired angles become entangled in complex multiplicative relationships. The addition of dynamic FAS positioning further increases this complexity by introducing time-varying spatial characteristics that must be jointly processed. More crucially, realistic multipath propagation environments create a fundamental challenge where line-of-sight (LoS) and NLoS multipath signal components become superimposed. This signal coupling not only degrades estimation accuracy but also increases computational complexity because traditional parameter estimation techniques struggle to disentangle the multiple propagation paths in the joint FAS-ARIS framework.
	
	To address these challenges, this work proposes a novel multi-stage estimation framework that strategically decouples the complex FAS-ARIS localization problem into manageable components. Through carefully designed pilot sequences and ARIS phase configurations, we enable effective separation of LoS and NLoS signals, followed by targeted parameter estimation for each component. The main contributions of this paper are summarized as follows:
	\begin{itemize}
		\item \textbf{System Design:} We propose an innovative FAS-ARIS-assisted SISO system that realizes 3D user localization without auxiliary channel information such as time-of-arrival (ToA) or frequency diversity, demonstrating adaptation to multipath environments.
		\item \textbf{Signal Decoupling:} We develop a novel pilot sequence and ARIS phase configuration design that effectively separates LoS and NLoS signal components, enabling independent estimation of AoA parameters at both the base station (BS) and ARIS.
		\item \textbf{Parameter Estimation:} We present a multi-stage estimation scheme that sequentially estimates the direct path angles using multiple signal classification (MUSIC) and recovers the cascaded channel parameters via maximum likelihood estimation (MLE) with interior-point refinement.
		\item \textbf{Localization Algorithm:} We develop a geometric localization approach using least-squares (LS) estimation to determine the UE's 3D position from the estimated AoA parameters.
		\item \textbf{Performance Analysis:} We derive Cram\'{a}r-Rao bounds (CRB) for channel parameters and position error bounds as theoretical benchmarks, providing fundamental performance limits for FAS-ARIS localization systems.
		\item \textbf{Simulation Validation:} We conduct extensive simulations studying the impact of key system parameters including transmit power, ARIS amplification factors, FAS spatial DoF, and multipath scatterers, confirming the effectiveness of the proposed framework.
	\end{itemize}
	
	Organization---The remainder of this paper is organized as follows. Section \ref{sec:2} presents the FAS-ARIS system model and channel formulation. Then Section \ref{sec:3} details the channel parameter estimation framework including signal decoupling and the multi-stage estimation algorithm. Section \ref{sec:4} describes the geometric 3D localization approach. Section \ref{sec:5} derives the theoretical performance bounds. Section \ref{sec:results} provides comprehensive simulation results and performance analysis. Finally, Section \ref{sec:conclude} concludes the paper.

	\begin{table}[!t]
		\centering
		\caption{\color{blue}SYMBOLS OF KEY VARIABLES}
		\renewcommand{\arraystretch}{2}
		  {\color{blue}
		\begin{tabular}{cc}
			\hline
			Symbol  & Definition           \\
			\hline
			$M_R$ & elements of ARIS \\
			$N$ &   number of FAS moves\\
			$\mathbf{p}_{\mathrm{R(B, U)}}$ & the center of ARIS (BS, UE) position   \\
			$\mathbf{p}_{{\mathrm{R(B)}, m(n))}}$ & $m$($n$)-th ARIS (BS) position   \\ 
			$L_\textrm{S1(S2,S3)}$ & number of scatterers of UE-ARIS (ARIS-BS, UE-BS) link   \\
			${\mathbf{p}}_{L_{\textrm{S}i,j}}$ & the position of the $j$-th scatter in the $Si$-th path \\
			${\bm{\theta}}_{\textrm{UR},L_\textrm{S1}}^{j}$ & AoAs at the ARIS from UE  \\
			${\bm{\theta}}_{\textrm{UB},L_\textrm{S3}}^{j}$ & AoAs at the BS from UE \\
			${\bm{\theta}}_{\textrm{RB},L_\textrm{S2}}^{j}$ & AoAs at the BS from ARIS  \\
			${\bm{\theta}}_{\textrm{BR},L_\textrm{S2}}^{j}$ & AoDs at the ARIS from BS \\
			$\rho_\textrm{UR}^i$ & UE-ARIS path gain \\
			$\rho_\textrm{UB}^i$ & UE-BS path gain \\
			$\rho_\textrm{BR(RB)}^i$ & ARIS-BS path gain \\
			\hline
			
		\end{tabular}}\label{table1}
	\end{table}
	
	Notations---Matrices and vectors are denoted by boldfaced uppercase and lowercase letters, respectively. The notations $ \otimes $  and $ \odot $  represent Kronecker product, Hadamard product, respectively. The superscripts ${\textbf{A}^{{T}}}$, ${\textbf{A}^*}$, ${\textbf{A}^{\text{H}}}$, ${\textbf{A}^\dag}$ stand for transpose, conjugate, conjugate transpose, and pseudoinverse of $\textbf{A}$, respectively. Moreover, $\Re $ denotes taking the real part while $\Im $ denotes taking the imaginary part. Furthermore, $\left|  \cdot  \right|$ and $\left\|  \cdot  \right\|$ indicate modulus, and $l_2$-norm, respectively. Also, ${\text{\textrm{diag}}}\left(  \cdot  \right)$ is the \textrm{diag}onal of a vector: ${\text{Vec}}\left(  \cdot  \right)$ and ${\text{Tr}}\left(  \cdot  \right)$ are the vectorization and trace of a matrix, respectively.
	
	\section{System and Signal Models}\label{sec:2}
	\subsection{System Model}
	\begin{figure*}[ht]
		\centering
		\centerline{\includegraphics[width=11cm]{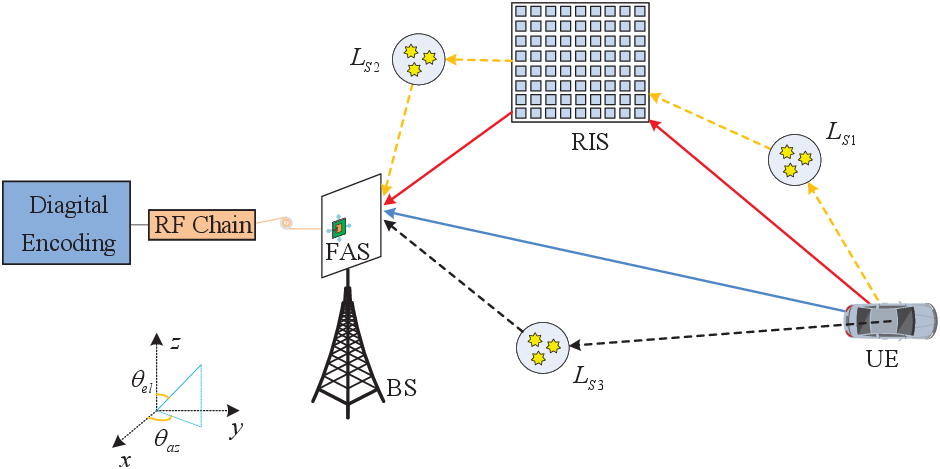}}
		\caption{The system model of FAS-ARIS uplink communication and localization system.}\label{fig1}
	\end{figure*}
	
	As shown in Fig.~\ref{fig1}, we consider an uplink communication system designed for 3D localization, where a user equipment (UE) is assisted by an ARIS  and an FAS-equipped BS. The system consists of three main components. First, a single-antenna UE, whose position ${\mathbf{p}}_\textrm{U} \in {\mathbb{R}^3}$ is unknown and to be estimated. Second, a BS equipped with an FAS, located at a known position ${{\mathbf{p}}_\textrm{B}} \in {\mathbb{R}^3}$. Third, an ARIS, situated at a known center position ${{\mathbf{p}}_\textrm{R}} \in {\mathbb{R}^3}$, which comprises ${M_\textrm{R}} = {M_\textrm{x}}{M_\textrm{z}}$ elements, with the $i$-th element's location denoted by ${{\mathbf{p}}_{R,i}}$ for $i \in \left\{ {1,2, \dots ,{M_\textrm{R}}} \right\}$. The localization process is based on an uplink transmission, where the UE sends $T$ pilot signals to the BS. These signals are received via both the direct UE-BS path and the reflective UE-RIS-BS path. 
	
	Furthermore, the geometric arrangement of the system is defined within a Cartesian coordinate system originating at the center of the BS {(the center of the virtual array is formed by FAS)}. The ARIS is modeled as a uniform planar array (UPA) parallel to the $y-o-z$ plane, with an element spacing of $d = \lambda/2$, where $\lambda$ is the signal wavelength. At the BS, the FAS consists of a single antenna with capability of switching its position within a 2D plane parallel to the $x-o-z$ plane. Over the observation period, the FAS occupies $N$ discrete positions, denoted by ${{\mathbf{p}}_\textrm{B,n}} = {[{x_\textrm{n}},0,{z_\textrm{n}}]^{{T}}} \in {\mathbf{C}}$ for $n \in \left\{ {1,2, \dots ,N} \right\}$, all contained within a square region $\mathbf{C}$ of size $A\lambda  \times A\lambda$. To capture a realistic propagation environment, our model incorporates multipath effects from a total of $L_\textrm{S} = L_\textrm{S1} + L_\textrm{S2} + L_\textrm{S3}$ unknown scatterers. These are distributed with $L_\textrm{S1}$ scatterers present in the UE-RIS path, $L_\textrm{S2}$ in the ARIS-BS path, and $L_\textrm{S3}$ in the UE-BS path. The position of the $j$-th scatterer is denoted by ${\mathbf{p}}_{L_{\textrm{S}i},j}$, where the index $i \in \{1,2,3\}$ corresponds to one of these three paths.

	\subsection{Channel Model}
	The channel model is developed under a standard far-field propagation assumption. A key simplification is that while the FAS antenna switches across its designated positions, its total displacement is assumed to be negligible compared to the distances between the UE, ARIS, and BS. This allows us to treat the AoAs, angles of departure (AoDs), and complex path gains as constant over the observation interval. Consequently, the channel variations are captured solely by the phase shifts corresponding to the different FAS positions. Based on these assumptions, the far-field steering vectors for the ARIS and the virtual array formed by the FAS are defined as 
{
	\begin{align}\label{eq1}
		&{{\mathbf{a}}_\textrm{R}}(\bm{\theta}) = \big[ e^{-j\frac{2\pi}{\lambda}\, \mathbf{p}_\textrm{R,1}^T\bm{k}(\bm{\theta})},\, \ldots,\, e^{-j\frac{2\pi}{\lambda}\, \mathbf{p}_{\textrm R,{\textrm M_\textrm{R}}}^T\bm{k}(\bm{\theta})} \big]^{{T}} 
		\in \mathbb{C}^{M_\textrm{R}}, \\
		&{{\mathbf{a}}_\textrm{B}}(\bm{\theta}) = \big[ e^{-j\frac{2\pi}{\lambda}\, \mathbf{p}_\textrm{B,1}^T\bm{k}(\bm{\theta})},\, \ldots,\, e^{-j\frac{2\pi}{\lambda}\, \mathbf{p}_\textrm{B,N}^T\bm{k}(\bm{\theta})} \big] ^{{T}}
		\in \mathbb{C}^{N}.
\end{align}}where ${\bm{\theta }} \triangleq {[{\theta_\textrm{el}},{\theta_\textrm{az}}]^{{T}}}$ denotes the angle pair, with ${\theta_\textrm{el}}$ and ${\theta_\textrm{az}}$ being the elevation and azimuth angles, respectively. The corresponding unit-norm direction vector ${\bm{k}}(\bm{\theta})$ is given by 
	\begin{equation}\label{eq2}
		{\bm{k}}(\bm{\theta}) = {\left[ \sin {\theta _\textrm{el}}\cos {\theta _\textrm{az}}, \sin {\theta _\textrm{el}}\sin {\theta _\textrm{az}}, \cos {\theta _\textrm{el}} \right]^{{T}}}.
	\end{equation} 
	
	Based on the positions of the system components, the specific angle pairs for each propagation path can be determined. For any two points, ${\mathbf p}_1$ and ${\mathbf p}_2$, the elevation and azimuth angles ($\theta_\textrm{el}$, $\theta_\textrm{az}$) of the directional vector pointing from ${\mathbf p}_2$ to ${\mathbf p}_1$ are computed as
	\begin{equation}\label{eq:angle_definitions}
		\begin{aligned}
			\theta_\textrm{el}({\mathbf p}_1, {\mathbf p}_2) &= \arccos\left( \frac{[{\mathbf p}_1-{\mathbf p}_2]_3}{\lVert {\mathbf p}_1-{\mathbf p}_2 \rVert} \right), \\
			\theta_\textrm{az}({\mathbf p}_1, {\mathbf p}_2) &= \operatorname{atan2}\left( \frac{[{\mathbf p}_1-{\mathbf p}_2]_2}{[{\mathbf p}_1-{\mathbf p}_2]_1} \right).
		\end{aligned}
	\end{equation}
	
	Using this generalized definition, we specify the angle pairs for all LoS and multipath components in the system. For the direct (LoS) {and reflected (NLOS)} paths, the angle pairs are denoted as
	\begin{itemize}
		\item ${\bm{\theta}}_\textrm{UR}$: AoA at the ARIS from the UE, computed with $({\mathbf p}_\textrm{U}, {\mathbf p}_\textrm{R})$.
		\item ${\bm{\theta}}_\textrm{UB}$: AoA at the BS from the UE, computed with $({\mathbf p}_\textrm{U}, {\mathbf p}_\textrm{B})$.
		{\item${\bm{\theta}}_\textrm{RB}$: AoA at the BS from the ARIS, computed with $({\mathbf p}_\textrm{R}, {\mathbf p}_\textrm{B})$.}
		{\item${\bm{\theta}}_\textrm{BR}$: AoD at the ARIS from the BS, computed with $({\mathbf p}_\textrm{B}, {\mathbf p}_\textrm{R})$.}
	\end{itemize}
	
	For the scattered paths, the angles are defined between a scatterer and an endpoint (BS or ARIS). For example, ${\bm{\theta}}_{\textrm{UR},L_\textrm{S1}}^{j}$ represents the AoA at the ARIS from the $j$-th scatterer in the UE-RIS link, calculated using $({\mathbf p}_{L_\textrm{S1}}^{\,j}, {\mathbf p}_\textrm{R})$. The angles for other scattered paths (${\bm{\theta}}_{\textrm{UB},L_\textrm{S3}}^{j}$, ${\bm{\theta}}_{\textrm{RB},L_\textrm{S2}}^{j}$, and ${\bm{\theta}}_{\textrm{BR},L_\textrm{S2}}^{j}$) are defined similarly. Since the UE and scatterers are modeled as single points, the angles from the UE to the scatterers are not considered.
	
	
	The channel between the UE and the ARIS, denoted by ${\mathbf{H}}_\textrm{UR} \in \mathbb{C}^{M_\textrm{R}}$, comprises a LoS path and $L_\textrm{S1}$ scattered paths. The channel response at the $m$-th ARIS element is given by
	\begin{align}\label{eq:H_UR}
		\mathbf{H}_\textrm{UR}(m) = \sum_{i=0}^{L_\textrm{S1}} \rho_\textrm{UR}^i\, e^{-j\frac{2\pi}{\lambda}\, \mathbf{p}_{\textrm R,m}^\textrm{T}\bm{k}(\bm{\theta}_\textrm{UR}^i)},
	\end{align}
	for $m = 1, \dots, M_\textrm{R}$, where $i=0$ corresponds to the LoS path, $\rho_\textrm{UR}^i$ is the complex path gain, and ${\bm{\theta}}_\textrm{UR}^i$ is the AoA at the ARIS for the $i$-th path.
	
	Similarly, the channel between the UE and the BS is captured at the $N$ discrete positions of the FAS. The channel response for the $n$-th FAS position, $\mathbf{H}_\textrm{UB}(n)$, consists of an LoS path and $L_\textrm{S3}$ scattered paths, expressed as
	\begin{align}\label{eq:H_UB}
		\mathbf{H}_\textrm{UB}(n) = \sum_{i=0}^{L_\textrm{S3}} \rho_\textrm{UB}^i\, e^{-j\frac{2\pi}{\lambda}\, \mathbf{p}_{\textrm B,n}^\textrm{T}\bm{k}(\bm{\theta}_\textrm{UB}^i)},
	\end{align}
	for $n=1, \dots, N$, where $\rho_\textrm{UB}^i$ and ${\bm{\theta}}_\textrm{UB}^i$ are the complex path gain and the AoA at the BS for the $i$-th path, respectively.
	
	The channel from the ARIS to the BS, ${\mathbf{H}}_\textrm{RB} \in \mathbb{C}^{N \times M_\textrm{R}}$, is modeled similarly. It consists of a direct path and $L_\textrm{S2}$ scattered paths. The channel coefficient from the $m$-th ARIS element to the $n$-th FAS position is
	\begin{align}\label{eq:H_RB}
		\mathbf{H}_\textrm{RB}(n,m) = \sum_{i=0}^{L_\textrm{S2}} \rho_\textrm{RB}^i e^{-j\frac{2\pi}{\lambda}(\mathbf{p}_{\textrm B,n}^\textrm{T}\bm{k}(\bm{\theta}_\textrm{BR}^i) + \mathbf{p}_{\textrm R,m}^\textrm{T}\bm{k}(\bm{\theta}_\textrm{RB}^i))},
	\end{align}
	where $\rho_\textrm{RB}^i$ is the complex gain, while ${\bm{\theta}}_\textrm{BR}^i$ and ${\bm{\theta}}_\textrm{RB}^i$ are the AoD from the ARIS and AoA at the BS for the $i$-th path, respectively.
	
	\subsection{Signal Model}
	During the uplink transmission, the UE transmits a sequence of $T$ pilot symbols, denoted by $\mathbf{x} = [x_1, \dots, x_T]^\textrm{T}$ with transmit power $P_\textrm{U}$. The signal received at the $n$-th FAS position during the $t$-th time slot, $y_{n,t}$, is a superposition of the signal from the direct UE-BS link and the reflected UE-RIS-BS link.
	
	The ARIS introduces a phase shift vector ${\mathbf{w}}_t = [w_{t,1}, \dots, w_{t,M_\textrm{R}}]^\textrm{T}$ at each time slot $t$, where $|w_{t,m}| = p$ and $p > 1$ is the amplification factor. The cascaded UE-RIS-BS channel for the $n$-th FAS position at time $t$ is given by $\mathbf{H}_\textrm{RB}(n)^\textrm{T} \operatorname{diag}(\mathbf{w}_t) \mathbf{H}_\textrm{UR}$, where $\mathbf{H}_\textrm{UR} \in \mathbb{C}^{M_\textrm{R}}$ and $\mathbf{H}_\textrm{RB}(n) \in \mathbb{C}^{M_\textrm{R}}$ are the channel vectors defined by \eqref{eq:H_UR} and the $n$-th row of \eqref{eq:H_RB}, respectively.
	
	The total received signal $y_{n,t}$ is expressed as 
	{ \begin{align}\label{eq:y_nt}
		y_{n,t} &= \underbrace{\mathbf{H}_\textrm{UB}(n) x_t}_{\text{Direct Link (LoS)}} + \underbrace{\mathbf{H}_\textrm{RB}(n)^\textrm{T} \operatorname{diag}(\mathbf{w}_t) \mathbf{H}_\textrm{UR} x_t}_{\text{Reflected Link (NLoS)}} + z_{n,t},
	\end{align}}
	where $z_{n,t}$ is the total noise, comprising the thermal noise at the BS, $\mathbf{z}_{\textrm B} \sim \mathcal{CN}(0, \sigma_\textrm{B}^2)$, and the noise introduced by the ARIS, $\mathbf{z}_{\textrm R} \sim \mathcal{CN}(0, \sigma_\textrm{R}^2)$. By collecting the received signals for all $N$ positions and $T$ time slots, we can form the received signal matrix $\mathbf{Y} \in \mathbb{C}^{N \times T}$, where $[\mathbf{Y}]_{n,t} = y_{n,t}$. This is compactly written as 
	{{\begin{equation}\label{eq:Y_matrix}
				\mathbf{Y} = \mathbf{H}_\textrm{LoS} + \mathbf{H}_\textrm{NLoS} + \mathbf{Z},
		\end{equation}}{{where  the $(n,t)$-th element of the direct channel matrix $\mathbf{H}_\textrm{LoS} = [H_\textrm{UB}(1), \dots, H_\textrm{UB}(N)]^\textrm{T}\odot\mathbf{x}$}, and the $(n,t)$-th element of the reflected channel matrix $\mathbf{H}_\textrm{NLoS}=  \mathbf{H}_\textrm{RB}(n)^\textrm{T} \operatorname{diag}(\mathbf{w}_t) \mathbf{H}_\textrm{UR} x_t$.
	
	\subsection{Problem Formulation}
	The ultimate goal of this work is to estimate the unknown 3D position of the UE, ${{\mathbf{p}}_\textrm{U}}$, from the uplink signals collected by the FAS. The localization is performed by first estimating the essential geometric channel parameters from the received signal matrix $\mathbf{Y}$. Specifically, our methodology hinges on accurately determining the AoA of the direct path from the UE to the BS, ${\bm{\theta}}_\textrm{UB}$, and the AoA of the direct path from the UE to the ARIS, ${\bm{\theta}}_\textrm{UR}$.
	
	Based on the signal model developed in \eqref{eq:y_nt}, the estimation of these angles can be cast as a complex nonlinear LS problem. The objective is to find the set of angular parameters that minimizes the discrepancy between the observed measurements and the theoretical model. This problem can be formulated as 
	\begin{align}\label{eq:problem_formulation}
		\left\{ {\hat{\bm{\theta}}_\textrm{UB}^k, \hat{\bm{\phi}}^{i,j}} \right\}& = \mathop{\arg\min}_{\substack{\bm{\theta}_\textrm{UB}^k, \\ \bm{\phi}^{i,j}}} \sum_{t=1}^T \Bigg\| \mathbf{y}_{t} - \sum_{k=0}^{L_\textrm{S3}} \rho_\textrm{UB}^k \mathbf{a}_\textrm{B}(\bm{\theta}_\textrm{UB}^k) x_t \nonumber \\
		&- \sum_{i=0}^{L_\textrm{S2}}\sum_{j=0}^{L_\textrm{S1}} \rho_\textrm{URB}^{i,j} \mathbf{a}_\textrm{B}(\bm{\theta}_\textrm{BR}^i) \left(\mathbf{a}_\textrm{R}(\bm{\phi}^{i,j})^\textrm{T}\mathbf{w}_t\right) x_t \Bigg\|^2,
	\end{align}
	where $\mathbf{y}_t$ is the $t$-th column of $\mathbf{Y}$, $\rho_\textrm{URB}^{i,j} = \rho_\textrm{RB}^i \rho_\textrm{UR}^j$ is the composite channel gain, and $\bm{\phi}^{i,j}$ is the cascaded angle.
	
	Directly solving the optimization problem in \eqref{eq:problem_formulation} is computationally intractable, primarily due to the large number of parameters and the inherent coupling between the AoAs in the cascaded channel term. To overcome this, this paper proposes a practical multi-stage estimation framework. In the subsequent sections, we introduce a specific pilot sequence and ARIS phase configuration designed to decouple the direct (LoS) and reflected (NLoS) signal components. This decoupling enables the separate and more manageable estimation of ${\bm{\theta}}_\textrm{UB}$ from the LoS signal and the cascaded angles ${\bm{\phi}}^{i,j}$ from the NLoS signal, from which the desired ${\bm{\theta}}_\textrm{UR}$ can be recovered.
	
	It is important to clarify that while our model accommodates multipath effects, we do not estimate the positions of the scatterers. Instead, they are treated as sources of additional signal paths whose parameters are unknown. Our focus remains on robustly estimating the AoAs of the direct paths, as these are the crucial components for accurate UE localization.
	
	\section{Estimation of Channel Parameters}\label{sec:3}
	This section details the proposed channel parameter estimation framework. To overcome the intractability of jointly estimating all parameters in \eqref{eq:problem_formulation}, we introduce a specific design for the ARIS phase configuration and the UE's pilot sequence. This strategy allows for the effective separation of the LoS and NLoS signal components. Subsequently, the separated signals are processed independently to estimate the AoA at the BS from the LoS component and the AoA at the ARIS from the NLoS component.
	
	\subsection{RIS Phase and Pilot Design for Signal Decoupling}
	In the proposed system model, the received signal is a composite of LoS and NLoS components. While the ARIS can amplify the NLoS path to mitigate the multiplicative fading effect, mutual interference between the two paths complicates direct parameter estimation. To overcome this challenge and enable robust estimation, a specific design for the UE pilot sequence and the ARIS phase configuration is introduced to separate the LoS and NLoS signals.
	
	The total transmission duration, $T$, is set to be an even number. {Furthermore, this paper assumes that the channel matrix remains unchanged during the $T$ transmission duration}. The UE's pilot sequence is structured to transmit an identical block of symbols in two consecutive halves:
	\begin{equation}\label{eq21}
		{\mathbf{X}} = \left[ {\underbrace {{x_1},{x_2}, \dots ,{x_{T/2}}}_{{{\mathbf{X}}_1}},\underbrace {{x_1},{x_2}, \dots ,{x_{T/2}}}_{{{\mathbf{X}}_2}}} \right].
	\end{equation}
	
	Concurrently, the ARIS phase configuration is designed to be opposing in the two halves. The phase shifts in the second interval are the negative of those in the first:
	\begin{equation}\label{eq22}
		{{\mathbf{W}}_{}} = \left[ {\underbrace {{{\mathbf{w}}_1},{{\mathbf{w}}_2}, \dots {{\mathbf{w}}_{T/2}}}_{{{\mathbf{W}}_{{{\mathbf{T}}_1}}}},\underbrace { - {{\mathbf{w}}_1}, - {{\mathbf{w}}_2}, \dots  - {{\mathbf{w}}_{T/2}}}_{{{\mathbf{W}}_{{{\mathbf{T}}_2}}}}} \right].
	\end{equation}
	where ${{\mathbf{W}}_{{{\mathbf{T}}_1}}} = \left[ {{{\mathbf{w}}_1},{{\mathbf{w}}_2}, \dots {{\mathbf{w}}_{T/2}}} \right]$ and ${{\mathbf{W}}_{{{\mathbf{T}}_2}}} =  \left[ { - {{\mathbf{w}}_1}, - {{\mathbf{w}}_2}, \dots  - {{\mathbf{w}}_{T/2}}} \right]$ such that ${{\mathbf{W}}_{{{\mathbf{T}}_1}}} =  - {{\mathbf{W}}_{{{\mathbf{T}}_2}}}$. This specific configuration ensures that the NLoS path contribution is inverted in the second half of the transmission period, while the LoS path component remains unchanged.
	
	This design allows for the received signal matrix $\mathbf{Y}$ to be partitioned into two sub-matrices, ${{\mathbf{Y}}_{{\mathbf{T}}1}}$ and ${{\mathbf{Y}}_{{\mathbf{T}}2}}$, corresponding to the two transmission intervals. By simple algebraic manipulation, the LoS and NLoS components can be effectively decoupled. The LoS signal component is obtained by summing the two parts:
	\begin{equation}\label{eq23}
		\begin{aligned}
			\mathbf{Y}_{\text{LOS}} &= \frac{1}{2}(\mathbf{Y}_{{T}_1} + \mathbf{Y}_{{T}_2}) \\
			&= \begin{bmatrix}
				\mathbf{H}_\textrm{UB}^1 x_1 & \cdots & \mathbf{H}_\textrm{UB}^1 x_{T/2} \\
				\vdots & \ddots & \vdots \\
				\mathbf{H}_\textrm{UB}^N x_1 & \cdots & \mathbf{H}_\textrm{UB}^N x_{T/2}
			\end{bmatrix} + \mathbf{Z}_{\text{LOS}}.
		\end{aligned}
	\end{equation}
	
	Conversely, the NLoS signal component is isolated by taking their difference:
	\begin{equation}\label{eq24}
		\begin{aligned}
			\mathbf{Y}_{\text{NLOS}} &= \frac{1}{2}(\mathbf{Y}_{{T}_1} - \mathbf{Y}_{{T}_2}) \\
			&= \begin{bmatrix}
				\mathbf{H}_{RB,1}^{\text{tot}} x_1 & \cdots & \mathbf{H}_{RB,1}^{\text{tot}} x_{T/2} \\[2pt]
				\vdots & \ddots & \vdots \\[2pt]
				\mathbf{H}_{RB,N}^{\text{tot}} x_1 & \cdots & \mathbf{H}_{RB,N}^{\text{tot}} x_{T/2}
			\end{bmatrix} + \mathbf{Z}_{\text{NLOS}},
		\end{aligned}
	\end{equation}
	where $\mathbf{H}_{RB,n}^{\text{tot}} = \mathbf{H}_\textrm{RB}^{B,n}\mathbf{H}_\textrm{RB}^R\operatorname{\textrm{diag}}\{\mathbf{w}_t\}\mathbf{H}_\textrm{UR}$ for $n \in \{1,\ldots,N\}$ and $t \in \{1,\ldots,T/2\}$. This decoupling is fundamental to the subsequent parameter estimation stages.
	
	\subsection{Extracting the Direct Signal: AoA Estimation at the BS}\label{BS}
	With the direct path signal, ${\mathbf{Y}}_{{\text{LOS}}}$, successfully isolated in \eqref{eq23}, we can now proceed to estimate the AoAs at the BS, $\{{\bm{\theta }}_\textrm{UB}^i\}$. This task is a classic direction-of-arrival (DOA) estimation problem, for which the signal model can be expressed in the following compact matrix form:
	\begin{equation}\label{eq25}
		{{\mathbf{Y}}_{{\text{LOS}}}} = {\mathbf{AS}} + {{\mathbf{Z}}_{{\text{LOS}}}},
	\end{equation}
	where ${\mathbf{A}}$ is the steering matrix whose columns are the steering vectors for each arrival path, and ${\mathbf{S}}$ is the coefficient matrix containing the path gains and transmitted symbols.
	
	While various subspace-based methods can solve this problem \cite{34,35,36,37,38}, this work employs the well-established  MUSIC algorithm. The first step is to compute the sample covariance matrix of the received signal, given by
	\begin{equation}\label{eq26}
		\mathbf{R}{\text{ = }}{\frac{{{{\mathbf{Y}}_{{\text{LOS}}}}{\mathbf{Y}}_{{\text{LOS}}}^{\text{H}}}}{{T/2}}_{}} = \mathbf{A}{\mathbf{R}_{{{\mathbf{Y}}_{{\text{Los}}}}}}{\mathbf{A}^H} + {\sigma ^2}\mathbf{I}.
	\end{equation}
	
	By performing an eigenvalue decomposition on $\mathbf{R}$, we partition the eigenspace into a signal subspace, ${\mathbf{U}_\textrm{S}} \in {\mathbb{C}^{N \times B}}$, spanned by the eigenvectors corresponding to the $B$ largest eigenvalues, and a noise subspace, ${\mathbf{U}_\textrm{n}} \in {\mathbb{C}^{N \times (\frac{T}{2} - B)}}$, spanned by the remaining eigenvectors. Here, $B = {L_\textrm{S3}} + 1$ represents the total number of paths in the direct link.
	
	The core principle of MUSIC is that the steering vectors of the true arrival angles are orthogonal to the noise subspace. In an ideal, noise-free scenario,  we have
	\begin{equation}\label{eq27}
		{\mathbf{a}^H}\left( {\bm{\theta }} \right){\mathbf{U}_\textrm{n}} = 0.
	\end{equation}
	
	In practice, due to finite samples and noise, perfect orthogonality is not achieved. Therefore, the AoAs are estimated by searching for peaks in the MUSIC pseudo-spectrum, which is formulated as finding the angles that maximize
	\begin{equation}\label{eq28}
		{P_\text{MUSIC}} = \mathop {\arg \max }\limits_{{{\bm{\theta }}_\textrm{UB}}} \frac{1}{{{\bm{a}^H}\left( {{{\bm{\theta }}_\textrm{UB}}} \right){\mathbf{U}_\textrm{n}}\mathbf{U}_\textrm{n}^H\bm{a}\left( {{{\bm{\theta }}_\textrm{UB}}} \right)}}.
	\end{equation}
	
	The angles $\{{\mathbf{\theta }}_\textrm{UB}^i\}$ are found by identifying the $B$ largest peaks from a two-dimensional search of this pseudo-spectrum. To mitigate the grid-resolution limitation of the search, an interior-point method is then used to refine the initial estimates, which is given by
	\begin{equation}\label{eq29}
		\begin{aligned}
			\bm{\tilde \theta}_\textrm{UB}^i& = \mathop{\arg\min}\limits_{\bm{\theta}_\textrm{UB}^i} 
			\left\| \mathbf{Y}_{\text{LOS}}(:,T/2) - \bm{a}^H(\bm{\theta}_\textrm{UB}^i)x_{T/2} \right\| \\
			&\text{s.t.} \quad \bm{\theta}_\textrm{UB}^i \in [\bm{\theta}_\textrm{UB}^0 - \Delta\bm{\theta}, \bm{\theta}_\textrm{UB}^0 + \Delta\bm{\theta}].
		\end{aligned}
	\end{equation}
	
	Once the refined AoA estimates $\{{\bm{\tilde \theta }}_\textrm{UB}^i\}$ are obtained, the corresponding complex path gains can be recovered via a simple LS solution, which is formulated as
	\begin{equation}\label{eq30}
		\rho _\textrm{UB}^i = \frac{{{{\bm{a}}_{\text{B}}}{{({\bm{\tilde \theta }}_\textrm{UB}^i)}^\dag }{{\mathbf{Y}}_{{\text{LOS}}}}_{(:,T/2)}}}{{{x_{T/2}}}},i \in \left\{ {0, \cdots ,{L_\textrm{S3}}} \right\}.
	\end{equation}
	
	\begin{remark}
		Since this work does not aim to localize scatterers, only the parameters of the direct, no-scatterer path are required for the final localization step. This path can be reliably identified as the one corresponding to the strongest peak in the MUSIC spectrum, as signals reflected by scatterers are typically much weaker. Therefore, the search complexity can be significantly reduced by setting $B=1$ and focusing only on the most dominant path.
	\end{remark}

{
	\begin{remark}
	This work focuses on single-reflection paths since the primary objective is to estimate parameters for no-scatterers paths between nodes. Although multiple-reflection paths offer a more realistic representation of the propagation environment, they would introduce a substantial increase in the number of path paths. While these additional paths could enhance the strength of scattered paths, their power remains significantly weaker compared to that of no-scatterers paths. Consequently, reflection paths do not affect the angle screening process during estimation. Based on these reasons, this work only considers the case of single-reflection paths.
	\end{remark}
}
	\subsection{Unraveling the Reflected Path: AoA Estimation at the ARIS}
	Having extracted the direct path parameters, we now turn our attention to the more complex task of estimating the AoA at the ARIS from the isolated NLoS signal, ${{\mathbf{Y}}_{{\text{NLOS}}}}$. The primary challenge here lies in the cascaded nature of the UE-RIS-BS channel, where the desired angle, ${\mathbf{\theta }}_\textrm{UR}$, is entangled with the ARIS-BS channel parameters. To address this, we first introduce the concept of a ``cascaded angle" to simplify the signal model.
	
	This cascaded angle, denoted by $\bm{\phi}$, mathematically combines the AoA at the ARIS (${\bm{\theta}}_\textrm{UR}$) and the AoD from the ARIS (${\bm{\theta}}_\textrm{BR}$) into a single virtual angle parameter. The steering vector corresponding to this cascaded angle is defined as 
	\begin{equation}\label{eq31}
		\begin{aligned}
			\mathbf{a}^{{T}}(\bm{\phi}) &= \mathbf{a}_\textrm{R}^{{T}}(\bm{\theta}_\textrm{UR}) \odot \mathbf{a}_\textrm{R}(\bm{\theta}_\textrm{BR}) \\
			&= \left(\mathbf{a}_\textrm{R}(\bm{\theta}_\textrm{UR}) \otimes \mathbf{a}_c(\bm{\theta}_\textrm{UR})\right)^{{T}} \odot 
			\left(\mathbf{a}_\textrm{R}(\bm{\theta}_\textrm{BR}) \otimes \mathbf{a}_c(\bm{\theta}_\textrm{BR})\right) \\
			&= \left(\mathbf{a}_\textrm{R}(\bm{\phi}) \otimes \mathbf{a}_c(\bm{\phi})\right)^{{T}},
		\end{aligned}
	\end{equation}
	with its components defined as 
	\begin{equation}\label{eq32}
		\begin{aligned}
			\\
			[{\bf a}_{\textrm{R}}(\boldsymbol{\phi})]_{\textrm{n}}& =\\
			\exp&\left(-j\frac{2\pi nd}{\lambda}\left(\sin\theta_\textrm{UR}^\textrm{(el)}\cos\theta_\textrm{UR}^\textrm{(az)} + 
			\sin\theta_\textrm{BR}^\textrm{(el)}\cos\theta_\textrm{BR}^\textrm{(az)}\right)\right) \\
			&= \exp\left(-j\frac{2\pi nd}{\lambda}\sin\phi^\textrm{(el)}\cos\phi^\textrm{(az)}\right), \\ 
			[\mathbf{a}_c(\bm{\phi})]_\textrm{n} &= \exp\left(-j\frac{2\pi nd}{\lambda}\left(\cos\theta_\textrm{UR}^\textrm{(el)} + \cos\theta_\textrm{BR}^\textrm{(el)}\right)\right) \\
			&= \exp\left(-j\frac{2\pi nd}{\lambda}\cos\phi^\textrm{(el)}\right).
		\end{aligned}
	\end{equation}
	
	Due to multipath propagation, there exist $({L_\textrm{S1}} + 1) \times ({L_\textrm{S2}} + 1)$ such cascaded paths. Using this formulation, the NLoS signal from \eqref{eq24} can be expressed more explicitly. For the $t$-th time slot and $n$-th FAS position, the signal can be formulated as 
	{\begin{equation}\label{eq33}
		\begin{aligned}
			{{\mathbf{H}}_{{\text{NLOS}}}}_\textrm{n,t} &={\mathbf{H}}_\textrm{RB}^{B,n}{\mathbf{H}}_\textrm{RB}^R\operatorname{\textrm{diag}}\{ {{\mathbf{w}}_t}\} {{\mathbf{H}}_\textrm{UR}}{x_t} + {\mathbf{Z}}_\textrm{n,t}^{{\text{NLOS}}} \\ 
			&= \sum\limits_{k = 1}^{{M_\textrm{R}}} \sum\limits_{i = 0}^{{L_\textrm{S2}}} \sum\limits_{j = 0}^{{L_\textrm{S1}}} \rho _\textrm{RB}^i\rho _\textrm{UR}^j
			{e^{ - j2\pi ({{\mathbf{p}}_\textrm{B,n}^T}{\bm{k}}({\mathbf{\theta }}_{_\textrm{BR}}^i))/\lambda }} \\
			&\quad \times {e^{ - j2\pi ({{\mathbf{p}}_{R,k}^T}{\bm{k}}(\phi _{_{}}^{i,j}))/\lambda }}{{\mathbf{w}}_{t,k}}{x_t} + {\mathbf{Z}}_\textrm{n,t}^{{\text{NLOS}}},
		\end{aligned}
   \end{equation}}
	for a given time slot $t$, the signal vector across all FAS positions is written as
	\begin{equation}\label{eq34}
		\begin{aligned}
			&{{\mathbf{H}}_{{\text{NLOS}}}}_{(:,t)}\\
			&= \sum\limits_{i = 0}^{{L_\textrm{S2}}} {\sum\limits_{j = 0}^{{L_\textrm{S1}}} {\rho _\textrm{RB}^i\rho _\textrm{UR}^j{{\mathbf{a}}_\textrm{B}}^{{T}}\left( {{\mathbf{\theta }}_\textrm{RB}^i} \right){{\mathbf{a}}_\textrm{R}}\left( {{\bm{\phi} ^{i,j}}} \right){{\mathbf{w}}_t}{x_t}} }  + {\mathbf{Z}}_{(:,t)}^{{\text{NLOS}}}
		\end{aligned}
	\end{equation}
	
	Our estimation strategy proceeds in two stages. First, we estimate the AoAs at the BS for the NLoS paths, $\{{\bm{\theta }}_\textrm{RB}^i\}$. Then, we use these estimates to isolate and subsequently estimate the cascaded angles $\{{\bm{\phi} ^{i,j}}\}$.
	
	Similar to the procedure in Section \ref{BS}, the AoAs at the BS for the NLoS paths, $\{{\bm{\theta }}_\textrm{RB}^i\}$, can be estimated by applying the MUSIC algorithm to the covariance matrix of ${{\mathbf{Y}}_{{\text{NLOS}}}}$. The process is analogous to the one described for the LoS path and is omitted here for brevity.
	
	\begin{remark}
		A path matching challenge arises here. To address this, we process each estimated NLOS path at the BS individually. By removing the influence of a specific ${\bm{\theta }}_\textrm{RB}^i$, we can then estimate the corresponding cascaded angle ${\bm{\phi} ^{i,j}}$, thereby ensuring correct path association. As before, computational complexity can be significantly reduced by focusing only on the strongest path, which corresponds to the direct ARIS-BS link without scatterers.
	\end{remark} 
	
	With the estimates $\{{\bm{\tilde \theta }}_\textrm{RB}^i\}$ in hand, we can \textbf{\textit{peel off}}  their contribution from the signal by pre-multiplying with the pseudo-inverse of the corresponding steering vectors. This isolates the component dependent on the cascaded angles:
	\begin{equation}\label{eq35a}
		\begin{aligned}
			{{{\mathbf{\tilde H}}}_{{\text{NLOS}}}} &= \sum\limits_{i = 0}^{{L_\textrm{S2}}} {{{\mathbf{a}}_{\text{B}}}^\dag ({\bm{\tilde \theta }}_\textrm{RB}^i)} {{\mathbf{H}}_{{\text{NLOS}}}},
		\end{aligned}
	\end{equation}
	which results in a simplified signal model for each time slot $t$:
	\begin{equation}\label{eq35b}
		\begin{aligned}
			{\mathbf{\tilde H}}_{_{(:,t)}}^{{\text{NLOS}}} = \sum\limits_{i = 0}^{{L_\textrm{S2}}} \sum\limits_{j = 0}^{{L_\textrm{S1}}} 
			\rho _\textrm{RB}^i\rho _\textrm{UR}^j{{\mathbf{a}}_\textrm{R}}\left( {{\bm{\phi} ^{i,j}}} \right){{\mathbf{w}}_t}{x_t} + {\mathbf{Z}}_{(:,t)}^{{\text{NLOS}}}.
		\end{aligned}
	\end{equation}
	
	As such, we can estimate the cascaded angles by formulating a MLE problem and finding the parameters that best fit the isolated signal, which is given by
	\begin{align}\label{eq36}
		{\bm{\phi} ^{i,j}} = \mathop {\arg \min }\limits_{{\bm{\phi} ^{i,j}}} \left\| {{\mathbf{\tilde H}}_{_{(:,t)}}^{{\text{NLOS}}} - \sum\limits_{i = 0}^{{L_\textrm{S2}}} {\sum\limits_{j = 0}^{{L_\textrm{S1}}} {\rho _\textrm{RB}^i\rho _\textrm{UR}^j{{\mathbf{a}}_\textrm{R}}\left( {{\bm{\phi} ^{i,j}}} \right){{\mathbf{w}}_t}{x_t}} } } \right\|,
	\end{align}
	Building upon this, the estimates obtained from a 2D search are then refined using the interior-point method, written as 
	\begin{align}\label{eq37} 
		{{\tilde \phi }^{i,j}} =\mathop{\arg\min }\limits_{{{\tilde \phi }^{i,j}} \in [{{\tilde \phi }^{i,j}} - \Delta ,{{\tilde \phi }^{i,j}} + \Delta ]} &\Bigg\| {\mathbf{\tilde H}}_{_{(:,t)}}^{{\text{NLOS}}}- \nonumber\\
		&   \sum\limits_{i = 0}^{{L_\textrm{S2}}} {\sum\limits_{j = 0}^{{L_\textrm{S1}}} {\rho _\textrm{RB}^i\rho _\textrm{UR}^j{{\mathbf{a}}_\textrm{R}}\left( {{{\tilde \phi }^{i,j}}} \right){{\mathbf{w}}_t}{x_t}} }  \Bigg\|,  
	\end{align}
	where $\Delta $ denotes the refined grid offset. By identifying the strongest signal component, we obtain the cascaded angle for the direct UE-RIS-BS path, ${\tilde \phi ^{0,0}}$. Since the AoD from the ARIS to the BS, ${\theta _\textrm{BR}}$, is known from the system geometry, we can finally recover the desired AoA at the ARIS by rearranging the definitions in (\ref{eq31}) and (\ref{eq32}), which is given by
	\begin{align}\label{eq38} 
		\tilde \theta _\textrm{UR}^\textrm{(el)} &= \arccos(\cos [\tilde \phi _\textrm{(el)}^{0,0}] - (\cos [\theta _\textrm{BR}^\textrm{(el)}]) , \\
		\tilde \theta _\textrm{UR}^\textrm{(az)}& = \arccos(\frac{{\cos [\tilde \phi _\textrm{(el)}^{0,0}] - \sin [\theta _\textrm{BR}^\textrm{(el)}]\cos [\theta _\textrm{BR}^\textrm{(az)}]}}{{\sin [\tilde \theta _\textrm{UR}^\textrm{(el)}]}}).
	\end{align}
	Besides, the corresponding channel gains can then be estimated as 
	\begin{align}\label{eq39} 
		\rho _{URB}^{i,j} = \frac{{{{\mathbf{a}}_{\text{R}}}{{({{\tilde \phi }^{i,j}})}^\dag }{{\mathbf{Y}}_{{\text{NLOS}}}}_{(:,t)}}}{{{{\mathbf{w}}_t}{x_t}}}.  
	\end{align}
	
	{The overall channel parameter estimation and localization framework is summarized in Algorithm \ref{alg:FAS_ARIS_localization}.}
	
	\begin{algorithm}[H]
	\caption{	{FAS-ARIS 3D Localization Framework}}
		\label{alg:FAS_ARIS_localization}
		\begin{algorithmic}[1]
			\State {\textbf{Input:} Received signal matrix $\mathbf{Y}$, pilot sequence $\mathbf{X}$, ARIS phase configuration $\mathbf{W}$, BS position $\mathbf{p}_\textrm{B}$, ARIS position $\mathbf{p}_\textrm{R}$}
			\State {\textbf{Output:} Estimated UE 3D position $\hat{\mathbf{p}}_\textrm{U}$}
			\Statex
			\State {\textbf{Phase I: Signal Decoupling}}
			\State {Partition $\mathbf{Y}$ into $\mathbf{Y}_{{T}_1}$ and $\mathbf{Y}_{{T}_2}$ based on transmission intervals}
			\State {Compute LoS signal: $\mathbf{Y}_{\text{LOS}} = \frac{1}{2}(\mathbf{Y}_{{T}_1} + \mathbf{Y}_{{T}_2})$} \hfill \eqref{eq23}
			\State {Compute NLoS signal: $\mathbf{Y}_{\text{NLOS}} = \frac{1}{2}(\mathbf{Y}_{{T}_1} - \mathbf{Y}_{{T}_2})$} \hfill \eqref{eq24}
			\Statex
			\State {\textbf{Phase II: Direct Path AoA Estimation at BS}}
			\State {Compute covariance matrix: $\mathbf{R} = \frac{\mathbf{Y}_{\text{LOS}}\mathbf{Y}_{\text{LOS}}^{\text{H}}}{T/2}$} \hfill \eqref{eq26}
			\State {Perform eigenvalue decomposition: $\mathbf{R} = \mathbf{U}_\textrm{S}\mathbf{\Lambda}_\textrm{S}\mathbf{U}_\textrm{S}^H + \mathbf{U}_\textrm{n}\mathbf{\Lambda}_\textrm{n}\mathbf{U}_\textrm{n}^H$}
			\State {Apply MUSIC algorithm: $P_\text{MUSIC} = \arg\max_{\bm{\theta}_\textrm{UB}} \frac{1}{\mathbf{a}^H(\bm{\theta}_\textrm{UB})\mathbf{U}_\textrm{n}\mathbf{U}_\textrm{n}^H\mathbf{a}(\bm{\theta}_\textrm{UB})}$} \hfill \eqref{eq28}
			\State {Find dominant peak to get initial estimate $\bm{\theta}_\textrm{UB}^0$}
			\State {Refine estimate using interior-point method to obtain $\tilde{\bm{\theta}}_\textrm{UB}$} \hfill \eqref{eq29}
			\Statex
			\State {\textbf{Phase III: Reflected Path AoA Estimation at ARIS}}
			\State {\textit{Stage 1: Estimate NLoS AoA at BS}}
			\State {Apply MUSIC to $\mathbf{Y}_{\text{NLOS}}$ to find dominant NLoS AoA at BS: $\tilde{\bm{\theta}}_\textrm{RB}$}
			\State {\textit{Stage 2: Isolate cascaded angle}}
			\State {Remove $\tilde{\bm{\theta}}_\textrm{RB}$ contribution: $\mathbf{\tilde H}_{\text{NLOS}} = \mathbf{a}_{\text{B}}^\dag(\tilde{\bm{\theta}}_\textrm{RB}) \mathbf{Y}_{\text{NLOS}}$} \hfill \eqref{eq35a}
			\State {Estimate cascaded angle via MLE: $\bm{\phi}^{0,0} = \arg\min_{\bm{\phi}} \|\mathbf{\tilde H}_{\text{NLOS}} - \mathbf{a}_\textrm{R}(\bm{\phi})\mathbf{w}_t x_t\|$} \hfill \eqref{eq36}
			\State {Refine cascaded angle using interior-point method to obtain $\tilde{\bm{\phi}}^{0,0}$} \hfill \eqref{eq37}
			\State {\textit{Stage 3: Recover AoA at ARIS}}
			\State {Compute desired AoA: $\tilde{\bm{\theta}}_\textrm{UR} = f(\tilde{\bm{\phi}}^{0,0}, \bm{\theta}_\textrm{BR})$ using \eqref{eq38}}
			\Statex
			\State {\textbf{Phase IV: Geometric 3D Localization}}
			\State {Formulate LS problem: $f(\mathbf{p}_\textrm{U}) = \|(\mathbf{p}_\textrm{R} - \mathbf{p}_\textrm{U})^T\mathbf{K}_\textrm{R}(\mathbf{p}_\textrm{R} - \mathbf{p}_\textrm{U})\| + \|(\mathbf{p}_\textrm{B} - \mathbf{p}_\textrm{U})^T\mathbf{K}_\textrm{B}(\mathbf{p}_\textrm{B} - \mathbf{p}_\textrm{U})\|$} \hfill \eqref{eq42}
			\State {Solve for UE position: $\hat{\mathbf{p}}_\textrm{U} = (\mathbf{K}_\textrm{R} + \mathbf{K}_\textrm{B})^{-1}(\mathbf{K}_\textrm{R}\mathbf{p}_\textrm{R} + \mathbf{K}_\textrm{B}\mathbf{p}_\textrm{B})$} \hfill \eqref{eq44}
			\Statex
			\State {\textbf{Return} $\hat{\mathbf{p}}_\textrm{U}$}
		\end{algorithmic}
	\end{algorithm}
	
	\section{Geometric 3D Localization}\label{sec:4}
	With the direct-path AoAs at the BS (${\tilde{\bm{\theta}}}_\textrm{UB}$) and the ARIS (${\tilde{\bm{\theta}}}_\textrm{UR}$) successfully estimated, the final step is to determine the UE's 3D position, ${\mathbf{p}}_\textrm{U}$. This is achieved by leveraging the geometric relationship between the known positions of the BS and ARIS and the estimated AoAs. Based on the estimated angles, the UE's position can be expressed as 
	\begin{align}\label{eq40} 
		{{\mathbf{p}}_\textrm{U}} = {{\mathbf{p}}_\textrm{R}} + {{\mathbf{g}}_\textrm{R}}{\bm{k}}({{\mathbf{\theta }}_\textrm{UR}}),   \\
		{{\mathbf{p}}_\textrm{U}} = {{\mathbf{p}}_\textrm{B}} + {{\mathbf{g}}_\textrm{B}}{\bm{k}}({{\mathbf{\theta }}_\textrm{UB}}).  
	\end{align}
	Given the fact that ${\bm{k}}{({{\mathbf{\theta }}_\textrm{UR}})^T}{\bm{k}}({{\mathbf{\theta }}_\textrm{UR}}) = 1,\;{\bm{k}}{({{\mathbf{\theta }}_\textrm{UB}})^T}{\bm{k}}({{\mathbf{\theta }}_\textrm{UB}}) = 1$, therefore, we have
	\begin{align}\label{eq41} 
		{{\mathbf{g}}_\textrm{R}} = {\bm{k}}{({{\mathbf{\theta }}_\textrm{UR}})^T}({{\mathbf{p}}_\textrm{U}} - {{\mathbf{p}}_\textrm{R}}), \\    
		{{\mathbf{g}}_\textrm{B}} = {\bm{k}}{({{\mathbf{\theta }}_\textrm{UB}})^T}({{\mathbf{p}}_\textrm{U}} - {{\mathbf{p}}_\textrm{B}}).  
	\end{align}
	
	Due to estimation errors and noise, a LS problem is formulated to find the position ${\mathbf{p}}_\textrm{U}$ that best fits the geometry. The corresponding LS problem with respect to ${{\mathbf{p}}_\textrm{U}}$ can be formulated as 
	\begin{align}\label{eq42} 
		&f({{\mathbf{p}}_\textrm{U}}) \nonumber\\
		&= {\left\| {{{\mathbf{p}}_\textrm{R}} - {{\mathbf{p}}_\textrm{U}}{\text{ + }}{{\mathbf{g}}_\textrm{R}}{\bm{k}}({{\mathbf{\theta }}_\textrm{UR}})} \right\|^2} + {\left\| {{{\mathbf{p}}_\textrm{B}} - {{\mathbf{p}}_\textrm{U}}{\text{ + }}{{\mathbf{g}}_\textrm{B}}{\bm{k}}({{\mathbf{\theta }}_\textrm{UB}})} \right\|^2}  \nonumber\\
		&= {({{\mathbf{p}}_\textrm{R}} - {{\mathbf{p}}_\textrm{U}})^T}({{\mathbf{I}}_3} - {\bm{k}}({{\mathbf{\theta }}_\textrm{UR}}){\bm{k}}{({{\mathbf{\theta }}_\textrm{UR}})^T})({{\mathbf{p}}_\textrm{R}} - {{\mathbf{p}}_\textrm{U}}) \nonumber\\
		&+ {({{\mathbf{p}}_\textrm{B}} - {{\mathbf{p}}_\textrm{U}})^T}({{\mathbf{I}}_3} - {\bm{k}}({{\mathbf{\theta }}_\textrm{UB}}){\bm{k}}{({{\mathbf{\theta }}_\textrm{UB}})^T})({{\mathbf{p}}_\textrm{B}} - {{\mathbf{p}}_\textrm{U}}) \nonumber\\
		&\triangleq {({{\mathbf{p}}_\textrm{R}} - {{\mathbf{p}}_\textrm{U}})^T}{{\bm{k}}_\textrm{R}}({{\mathbf{p}}_\textrm{R}} - {{\mathbf{p}}_\textrm{U}}) + {({{\mathbf{p}}_\textrm{B}} - {{\mathbf{p}}_\textrm{U}})^T}{{\bm{k}}_{\text{B}}}({{\mathbf{p}}_{\text{B}}} - {{\mathbf{p}}_\textrm{U}}) ,
	\end{align}
	where  ${{\bm{k}}_{\text{R}}} \triangleq {{\mathbf{I}}_3} - {\bm{k}}({{\mathbf{\theta }}_\textrm{UR}}){\bm{k}}{({{\mathbf{\theta }}_\textrm{UR}})^T},{{\bm{k}}_{\text{B}}} \triangleq {{\mathbf{I}}_3} - {\bm{k}}({{\mathbf{\theta }}_{U{\text{B}}}}){\bm{k}}{({{\mathbf{\theta }}_\textrm{UB}})^T}$.
	
	To obtain the estimate of ${{\mathbf{p}}_\textrm{U}}$, we find the minimum of (\ref{eq42}) by setting its derivative to zero, which is formulated as
	\begin{equation}\label{eq43}
		\frac{{\partial f({{\mathbf{p}}_\textrm{U}})}}{{\partial {{\mathbf{p}}_\textrm{U}}}} = 0.
	\end{equation}
	This yields the final closed-form estimate of the UE's location, which is given by
	\begin{equation}\label{eq44}
		{{\mathbf{p}}_\textrm{U}} = {({{\bm{k}}_\textrm{R}} + {{\bm{k}}_\textrm{B}})^{ - 1}}({{\bm{k}}_\textrm{R}}{{\mathbf{p}}_\textrm{R}} + {{\bm{k}}_\textrm{B}}{{\mathbf{p}}_\textrm{B}}).
	\end{equation}
	
	\section{Performance Bounds}\label{sec:5}
	In this section, we derive the CRB to establish a theoretical performance benchmark for the estimators proposed in Sections \ref{sec:3} and \ref{sec:4}. We first derive the CRB for the channel parameters and then transform it to obtain the Position Error Bound (PEB) for the UE localization.
	
	\subsection{CRB for Channel Parameters}
	The vector of all unknown real channel parameters for the direct paths is defined as ${\bm{\gamma }} = {[\Re ({\rho _\textrm{UB}}),\Im ({\rho _\textrm{UB}}),\Re ({\rho _{URB}}),\Im ({\rho _{URB}}),{\mathbf{\theta }}_\textrm{UB}^\textrm{el},{\mathbf{\theta }}_\textrm{UB}^\textrm{az},{\mathbf{\theta }}_\textrm{UR}^\textrm{el},{\mathbf{\theta }}_\textrm{UR}^\textrm{az}]^T} \in {\mathbb{R}^8}$. The Fisher Information Matrix (FIM) for this parameter vector, ${\mathbf{F}}({\bm{\gamma }})$, provides the basis for the CRB:
	\begin{equation}\label{eq45}
		{\mathbf{F}}({\bm{\gamma }}) = \sum\limits_{n = 1}^N {\sum\limits_{t = 1}^T {\Re \left\{ {{{(\frac{{\partial {{\bm{\mu }}_\textrm{n,t}}}}{{\partial {\bm{\gamma }}}})}^H}{{({\mathbf{C}_\textrm{R}} + {\mathbf{C}_\textrm{B}})}^{ - 1}}(\frac{{\partial {{\bm{\mu }}_\textrm{n,t}}}}{{\partial {\bm{\gamma }}}})} \right\}} }
	\end{equation}
	where ${\mathbf{C}_\textrm{B}} = \sigma _\textrm{B}^2{\mathbf{I}}$ is the thermal noise covariance at the BS, ${\mathbf{C}_\textrm{R}} = \sigma _\textrm{R}^2p_\textrm{R}^2{{\mathbf{H}}_\textrm{RB}}{({{\mathbf{H}}_\textrm{RB}})^{\text{H}}}$ is the covariance of the noise from the ARIS, and ${\bm{\mu }}$ is the noise-free received signal:
	{{\begin{align}\label{eq46}
		{\bm{\mu}}_\textrm{n,t} &= \rho_\textrm{UB} \cdot e^{-j\frac{2\pi}{\lambda} \mathbf{p}_\textrm{B,n}^T \bm{k}({\bm{\theta}}_\textrm{UB})}   x_t +  \nonumber \\
		& \rho_\textrm{URB}  \sum_{i=1}^{M_\textrm{R}} e^{-j\frac{2\pi}{\lambda} \left( \mathbf{p}_\textrm{B,n}^T \bm{k}({\bm{\theta}}_\textrm{BR}) + \mathbf{p}_\textrm{R,1}^T \bm{k}({\bm{\theta}}_\textrm{RB}) + \mathbf{p}_\textrm{R,1}^T \bm{k}({\bm{\theta}}_\textrm{UR}) \right)}  \mathbf{w}_{t,i}  x_t.
	\end{align}}
	
	The specific derivations for the FIM are detailed in Appendix A. The CRB for any unbiased estimator of ${\bm{\gamma}}$ is given by the diagonal elements of ${{\mathbf{F}}^{ - 1}}({\bm{\gamma }})$.
	\begin{remark} The noise covariance ${C_\textrm{R}}$ is technically a function of the channel parameters. However, in line with common practice for CRB analysis, this dependency is treated as a second-order effect and is not differentiated when computing the FIM \cite{39}.
	\end{remark}
	
	\subsection{PEB}
	To evaluate the theoretical limit of localization accuracy, we transform the FIM from the channel parameter domain to the position parameter domain. The new parameter vector of interest is ${{\bm{\gamma }}_p} = {[\Re ({\rho _\textrm{UB}}),\Im ({\rho _\textrm{UB}}),\Re ({\rho _{URB}}),\Im ({\rho _{URB}}),{{\mathbf{p}}_\textrm{U}}^T]^T} \in {\mathbb{R}^7}$.
	
	The transformation is performed via the chain rule, using the Jacobian matrix ${\mathbf{J}} = \frac{{\partial {{\bm{\gamma }}^T}}}{{\partial {{\bm{\gamma }}_p}}} \in {\mathbb{R}^{8 \times 7}}$. The FIM for the position-based parameters is then given by
	\begin{equation}\label{eq47}
		{\mathbf{F}}({{\bm{\gamma }}_p}) = {\mathbf{JF}}({\bm{\gamma }}){{\mathbf{J}}^T}.
	\end{equation}
	The detailed derivation of the Jacobian matrix ${\mathbf{J}}$ is provided in Appendix B. Finally, the PEB is calculated from the inverse of this new FIM:
	\begin{equation}\label{eq48}
		{\text{PEB}} = \sqrt {{\rm Tr}\{ {{[{{\mathbf{F}}^{ - 1}}({{\bm{\gamma }}_p})]}_{5:7,5:7}}\} }.
	\end{equation}
	
	\section{Simulation Results}\label{sec:results}
	This section presents numerical results to evaluate the performance of the proposed channel estimation and localization algorithms. We first conduct simulations in an ideal scatterer-free environment to validate the fundamental performance of our method. Subsequently, we introduce scatterers to analyze the impact of multipath propagation on both estimation and localization accuracy.
	
	The simulations are configured for an outdoor far-field scenario. The channel gains for the BS-RIS, BS-UE, and ARIS-UE links follow the free-space path loss model: ${\rho _\textrm{RB}} = \frac{\lambda }{{4\pi \left\| {{{\mathbf{p}}_\textrm{R}} - {{\mathbf{p}}_\textrm{B}}} \right\|}}{e^{j{\alpha _\textrm{RB}}}}$, ${\rho _\textrm{UB}} = \frac{\lambda }{{4\pi \left\| {{{\mathbf{p}}_\textrm{U}} - {{\mathbf{p}}_\textrm{B}}} \right\|}}{e^{j{\alpha _\textrm{UB}}}}$, and ${\rho _\textrm{UR}} = \frac{\lambda }{{4\pi \left\| {{{\mathbf{p}}_\textrm{U}} - {{\mathbf{p}}_\textrm{R}}} \right\|}}{e^{j{\alpha _\textrm{UR}}}}$, with the cascaded gain being ${\rho _{URB}} = {\rho _\textrm{RB}}{\rho _\textrm{UR}}$. The phase components ${\alpha _\textrm{UB}}, {\alpha _\textrm{UR}}, {\alpha _\textrm{RB}}$ are drawn from a uniform distribution over $[0, 2\pi)$. The noise figure for both the receiver and the ARIS is set to 18 dB. Key simulation parameters are as follows: carrier frequency ${f_c} = 2.8$ GHz; number of FAS positions $N = 10 \times 10 = 100$; pilot sequence length $T = 100$; and a default power allocation factor for the ARIS $\varepsilon = 0.8$. The system geometry is defined by: BS position ${{\mathbf{p}}_\textrm{B}} = {[0,0,10]^{{T}}}$m, ARIS center ${{\mathbf{p}}_\textrm{R}} = {[ - 10,23.3,0.5]^T}$m, and the UE's true position ${{\mathbf{p}}_\textrm{U}} = {[3.5,26.7,0.7]^{{T}}}$m. The amplification factor of the ARIS, ${{\mathbf{p}}_\textrm{R}}$, is according to \cite{40}.
	
	{\subsection{RMSE versus $P$(dbm)}}
	Figs.~\ref{fig2a} to \ref{fig2c} illustrate the  RMSE  of the parameter estimates as a function of the UE's transmit power $P$ in a scatterer-free environment. Fig.~\ref{fig2a} shows the estimation performance for the AoA at the BS (${\bm{\theta }}_{{\text{UB}}}$). The RMSE curves for both elevation and azimuth angles closely track their corresponding CRBs, particularly for transmit powers above -10 {dbm}. This indicates that the proposed MUSIC-based estimation for the direct path is highly efficient after the signal decoupling stage.
	
	{
		To evaluate the accuracy of channel parameter estimation, the performance of algorithms in estimating ${\bm{\theta }}_{{\text{UB}}}$ is compared in Fig. \ref{figxiu1}, including estimation of signal parameters via rotational invariance techniques (ESPRIT), MUSIC, orthogonal matching pursuit (OMP), and least squares (LS). The results indicate that the MUSIC-based approach achieves the highest estimation accuracy, while ESPRIT attains comparable performance at higher signal power levels. In contrast, the accuracy of the OMP algorithm saturates early as $P$ increases. Although the LS method shows gradual improvement with rising signal power, it consistently underperforms compared to both MUSIC and ESPRIT. It is noteworthy that while the LS algorithm is known to achieve maximum likelihood performance under Gaussian noise conditions, this capability is compromised in the presence of thermal noise interference.}
	
	In contrast, {the LS-based method in} Fig.~\ref{fig2b} reveals a noticeable and persistent gap between the RMSE of the ARIS AoA estimate (${\bm{\theta }}_{{\text{UR}}}$) and its CRB across the entire power range. This performance gap is a direct consequence of our multi-stage estimation strategy for the cascaded channel. Unlike the direct estimation of ${\bm{\theta }}_{{\text{UB}}}$, the recovery of ${\bm{\theta }}_{{\text{UR}}}$ involves several sequential steps (estimating ${\bm{\theta }}_{{\text{RB}}}$, peeling it off, estimating the cascaded angle ${\bm{\phi}}$, and finally decoupling ${\bm{\theta }}_{{\text{UR}}}$). Errors introduced in each step propagate through the chain, leading to a cumulative performance loss compared to the theoretical bound of a joint estimator.
	
{
	Since cascaded channel parameters cannot be directly estimated using subspace methods, the estimation performance of algorithms for ${\bm{\theta }}_{{\text{UR}}}$ is compared in Fig. \ref{figxiu2}, including atomic norm minimization (ANM), LS, OMP, and nuclear norm minimization (NUC). The results show that as $P$ increases, both the OMP algorithm and nuclear norm-based method exhibit early saturation in estimation accuracy. The ANM-based method achieves accuracy comparable to that of the LS algorithm. Since the influence of ${\bm{\theta }}_{{\text{RB}}}$ has been eliminated and thermal noise is minimized to the greatest extent possible in the estimation, the performance of the LS algorithm is significantly improved.}
	
	This propagated error naturally affects the final localization accuracy, as shown in Fig.~\ref{fig2c}. The RMSE of the UE position estimate (${\mathbf{p}}_\textrm{U}$) exhibits a similar gap with respect to the PEB, which stems from the sub-optimality of the ${\bm{\theta }}_{{\text{UR}}}$ estimate and the use of a non-optimal LS method for the final geometric calculation.
	
	\begin{remark}
		The performance difference between the BS and ARIS AoA estimations provides a key insight into the proposed algorithm. It highlights the trade-off between problem tractability and statistical efficiency. While the multi-stage approach makes the complex estimation problem solvable, it introduces an inherent performance penalty, especially for the parameters of the reflected path. The estimator's efficiency is thus limited by the most challenging part of the estimation chain, disentangling the cascaded channel.
	\end{remark}
	
	\begin{figure}[htbp]
		\centering
		\includegraphics[width=1.1\columnwidth]{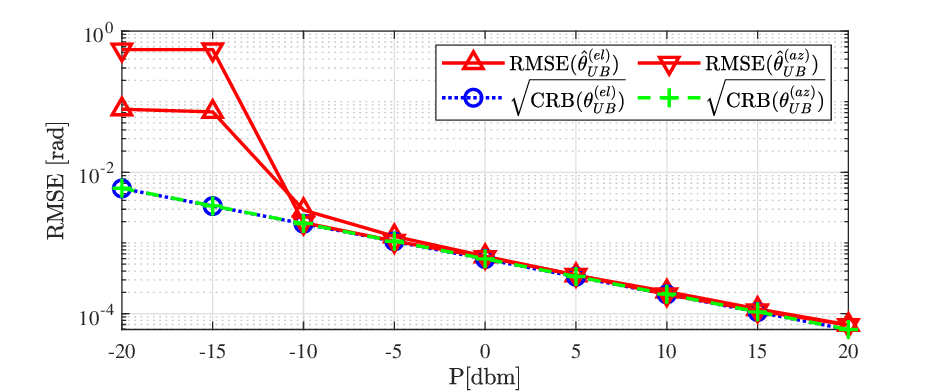}
		\caption{RMSE of ${{\bm{\theta }}_{{\text{UB}}}}$ versus $P$(dbm).}
		\label{fig2a}
	\end{figure}
	
	\begin{figure}[htbp]
		\centering
		\includegraphics[width=1.1\columnwidth]{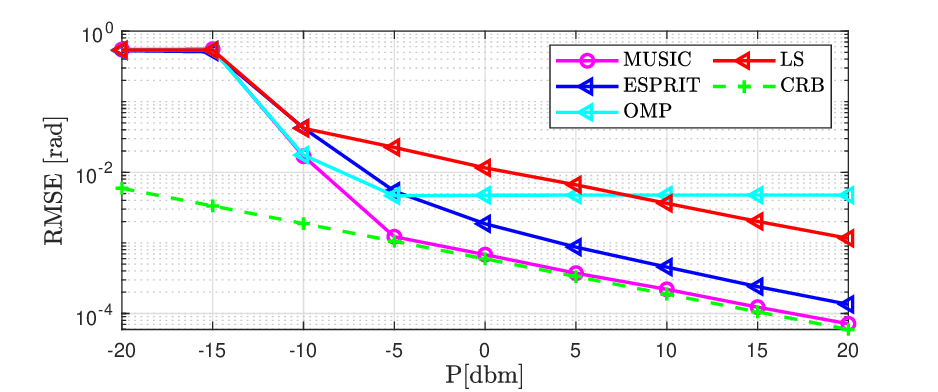}
		\caption{	{RMSEs of ${{\bm{\theta }}_{{\text{UB}}}}$ in different algorithms versus $P$(dbm).}}
		\label{figxiu1}
	\end{figure}
	
	\begin{figure}[htbp]
		\centering
		\includegraphics[width=1.1\columnwidth]{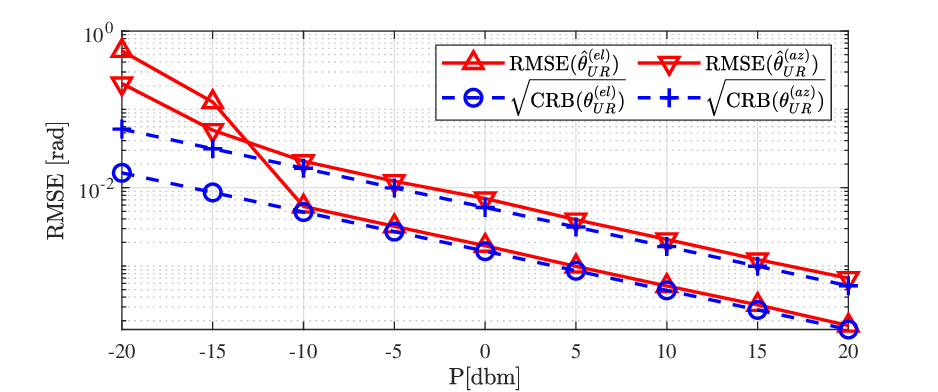}
		\caption{RMSE of ${{\bm{\theta }}_{{\text{UR}}}}$ versus $P$(dbm).}
		\label{fig2b}
	\end{figure}
	
	\begin{figure}[htbp]
		\centering
		\includegraphics[width=1.1\columnwidth]{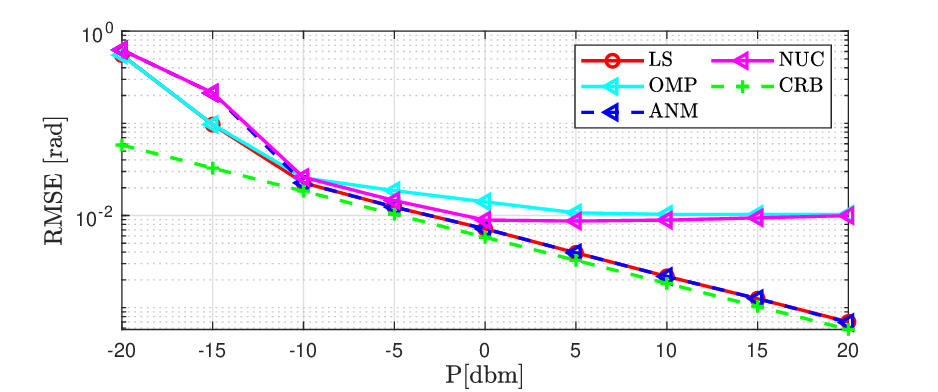}
		\caption{	{RMSEs of ${{\bm{\theta }}_{{\text{UR}}}}$ in different algorithms versus $P$(dbm).}}
		\label{figxiu2}
	\end{figure}
	
	\begin{figure}[htbp]
		\centering
		\includegraphics[width=1.1\columnwidth]{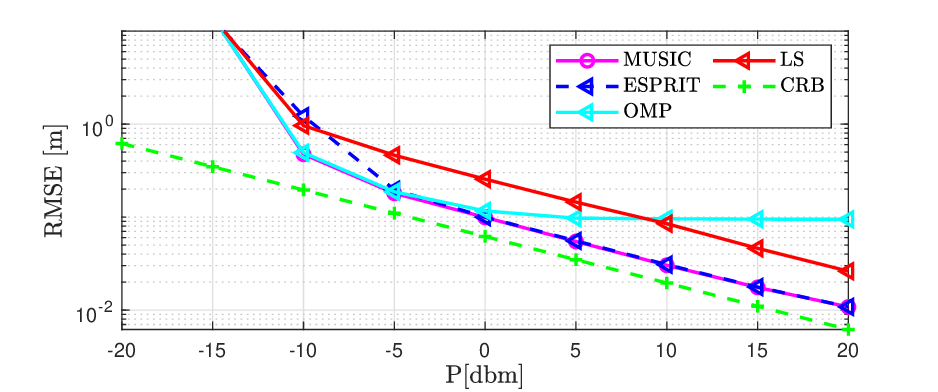}
		\caption{{RMSE of ${{\mathbf{p}}_\textrm{U}}$ versus $P$(dbm).}}
		\label{fig2c}
	\end{figure}
	
	\subsection{RMSE versus $\varepsilon$}
	This subsection analyzes the impact of the ARIS power allocation factor, $\varepsilon$, on estimation performance. The UE transmit power is fixed at $P=15$ {dbm}, while $\varepsilon$ is varied from 0.01 to 30. As shown in Fig.~\ref{fig3a}, the AoA estimation accuracy at the BS (${\bm{\theta }}_{{\text{UB}}}$) slightly degrades as $\varepsilon$ increases. This is because a larger $\varepsilon$ allocates more power to amplifying the NLoS signal and its associated ARIS-induced noise, which in turn reduces the relative SNR of the LoS path. {The trends exhibited in Fig.~\ref{figxiu3} by the various algorithms are consistent with those in Fig.~\ref{figxiu1}. }
	
	Conversely, Fig.~\ref{fig3b} demonstrates that the AoA estimation at the ARIS (${\bm{\theta }}_{{\text{UR}}}$) achieves its optimal performance for an $\varepsilon$ value between 0.1 and 0.5. This reveals a critical trade-off: $\varepsilon$ must be large enough to overcome the NLoS path's double fading but small enough to avoid excessive noise amplification. The U-shaped performance curve illustrates the existence of an optimal balancing point. {The trends exhibited in Fig.~\ref{figxiu4} by the various algorithms are also consistent with those in Fig.~\ref{figxiu2}.}
	
	The final localization error, depicted in Fig.~\ref{fig3c}, mirrors the performance trend of the ARIS AoA estimation, confirming that ${\bm{\theta }}_{{\text{UR}}}$ is the bottleneck for positioning accuracy. Notably, for $\varepsilon > 10$, the performance degrades catastrophically as the amplified noise overwhelms the system.
	
	\begin{remark}
		The power allocation factor $\varepsilon$ is a crucial system design parameter. The optimal choice of $\varepsilon$ depends on the relative path losses of the LoS and NLoS links. In practical systems, $\varepsilon$ should be dynamically adapted to the channel conditions to maintain optimal performance.
	\end{remark}
	
	\begin{figure}[htbp]
		\centering
		\includegraphics[width=1.1\columnwidth]{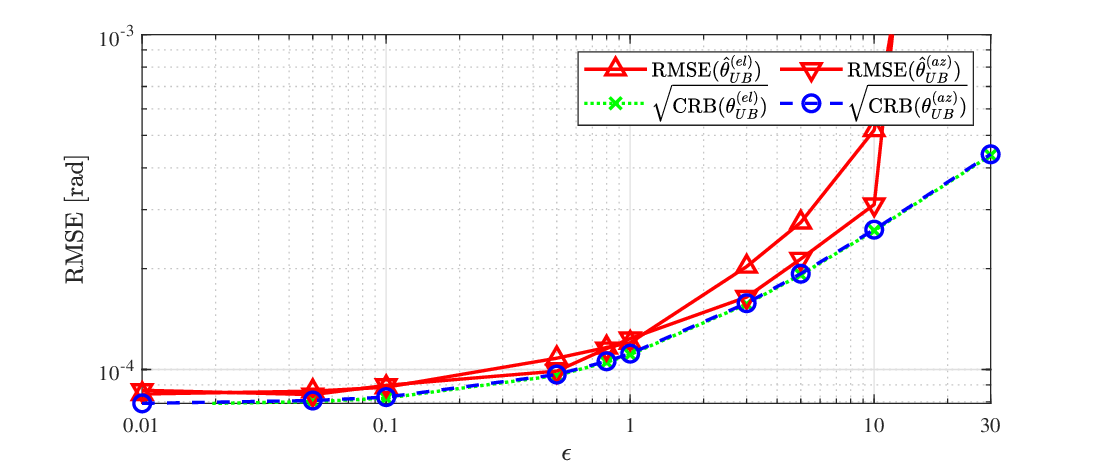}
		\caption{RMSE of ${{\bm{\theta }}_{{\text{UB}}}}$ versus $\varepsilon$.}
		\label{fig3a}
	\end{figure}
	
	\begin{figure}[htbp]
		\centering
		\includegraphics[width=1.1\columnwidth]{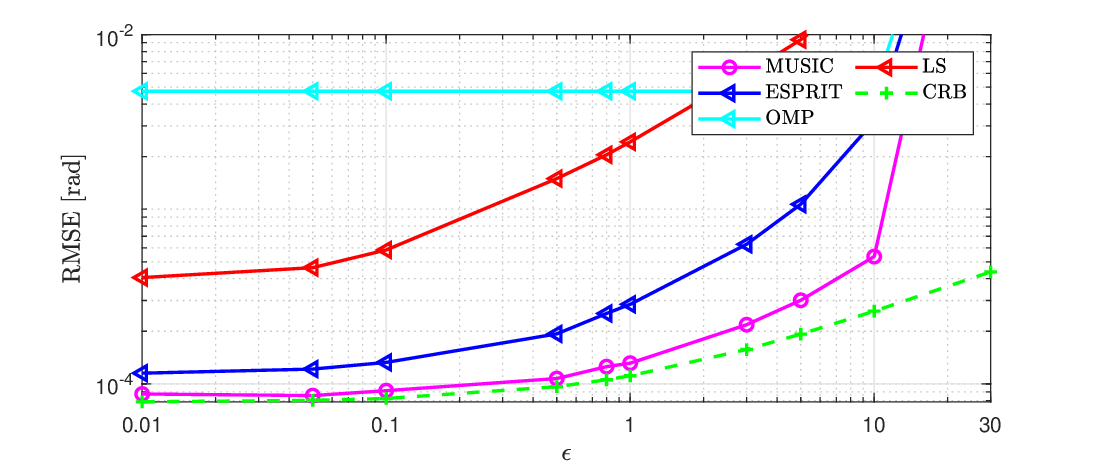}
		\caption{	{RMSEs of ${{\bm{\theta }}_{{\text{UB}}}}$ in different algorithms versus $\varepsilon$.}}
		\label{figxiu3}
	\end{figure}

	\begin{figure}[htbp]
		\centering
		\includegraphics[width=1.1\columnwidth]{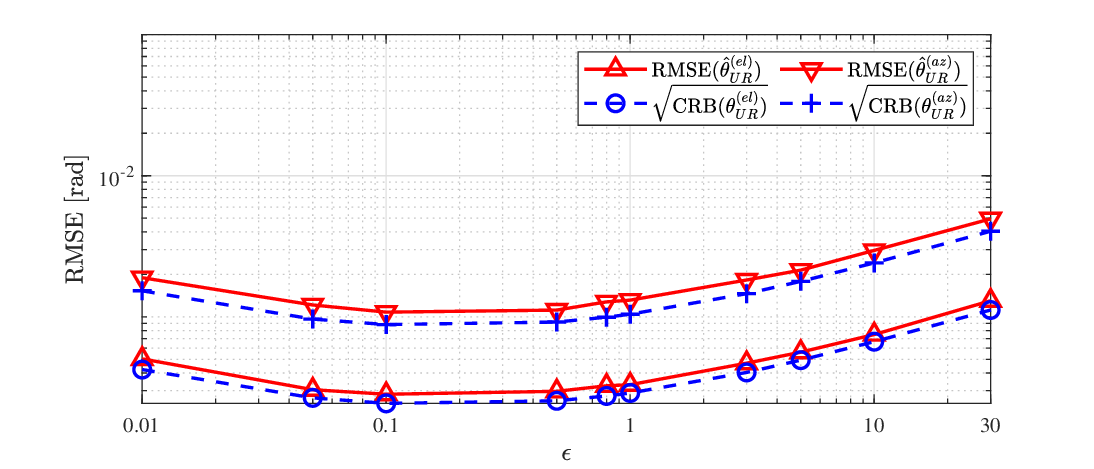}
		\caption{RMSE of ${{\bm{\theta }}_{{\text{UR}}}}$ versus $\varepsilon$.}
		\label{fig3b}
	\end{figure}
	
	\begin{figure}[htbp]
		\centering
		\includegraphics[width=1.1\columnwidth]{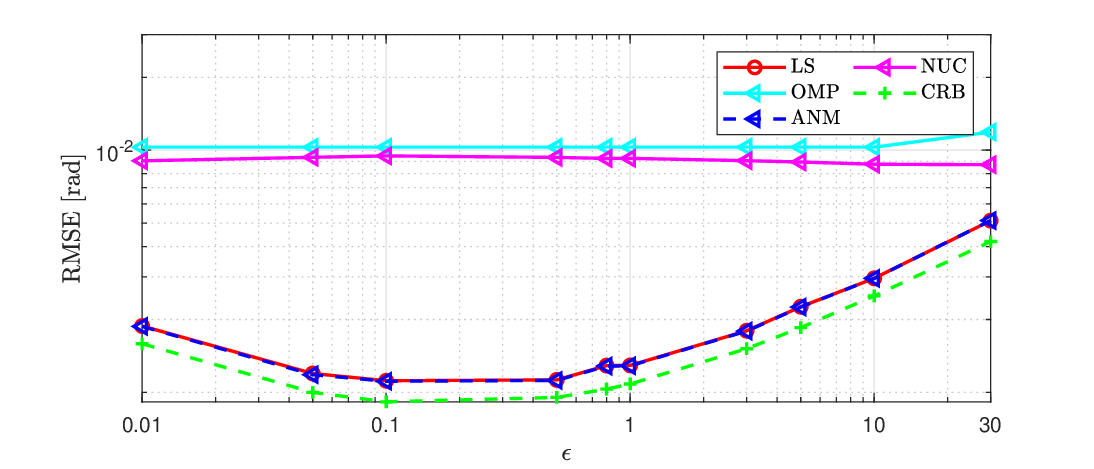}
		\caption{	{RMSEs of ${{\bm{\theta }}_{{\text{UR}}}}$ in different algorithms versus $\varepsilon$.}}
		\label{figxiu4}
	\end{figure}
	
	\begin{figure}[htbp]
		\centering
		\includegraphics[width=1.1\columnwidth]{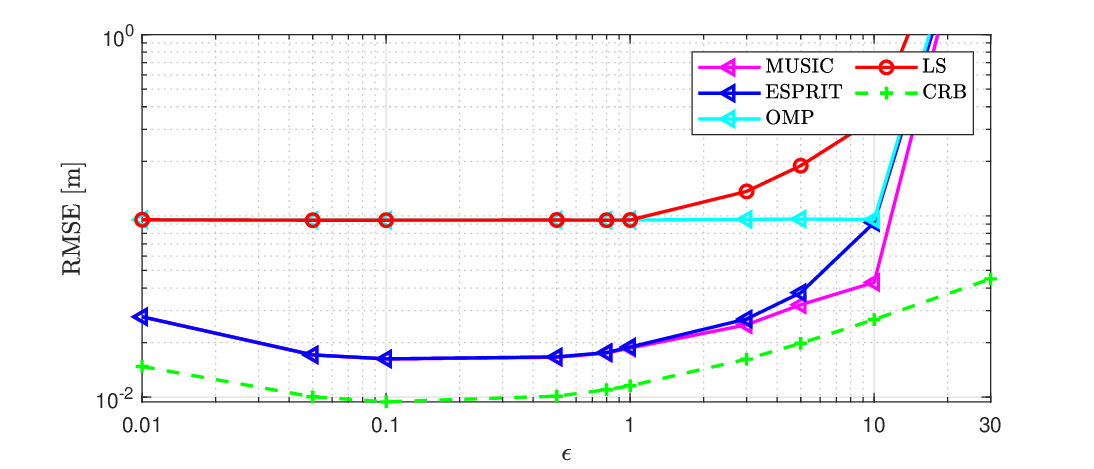}
		\caption{{RMSE of ${{\mathbf{p}}_\textrm{U}}$ versus $\varepsilon$.}}
		\label{fig3c}
	\end{figure}
	
	\subsection{RMSE versus number of FAS step sizes}
	This subsection investigates the impact of the spatial DoF provided by the FAS, varied by changing the number of antenna step sizes, $N$. The results are shown in Figs.~\ref{fig4a} to \ref{fig4c}, where ``EM" (Exhaustive Measurement) serves as a benchmark representing the performance with a continuous aperture. The figures consistently demonstrate that for all estimated parameters (${\bm{\theta }}_{{\text{UB}}}$, ${\bm{\theta }}_{{\text{UR}}}$, and ${\mathbf{p}}_\textrm{U}$), the estimation accuracy significantly improves as $N$ increases from 36 to 144. This is a direct result of the increased virtual aperture provided by the FAS with finer step sizes. A larger number of step sizes yields higher spatial resolution, which in turn leads to more precise angle estimation and, consequently, more accurate localization.
	
	\begin{remark}
		These results quantitatively validate the core motivation for integrating FAS into the localization system. The ability to dynamically sample the aperture with finer step sizes provides a substantial performance gain, effectively translating additional spatial DoF into enhanced estimation accuracy that approaches the theoretical benchmark.
	\end{remark}
	
	\begin{figure}[htbp]
		\centering
		\includegraphics[width=1.1\columnwidth]{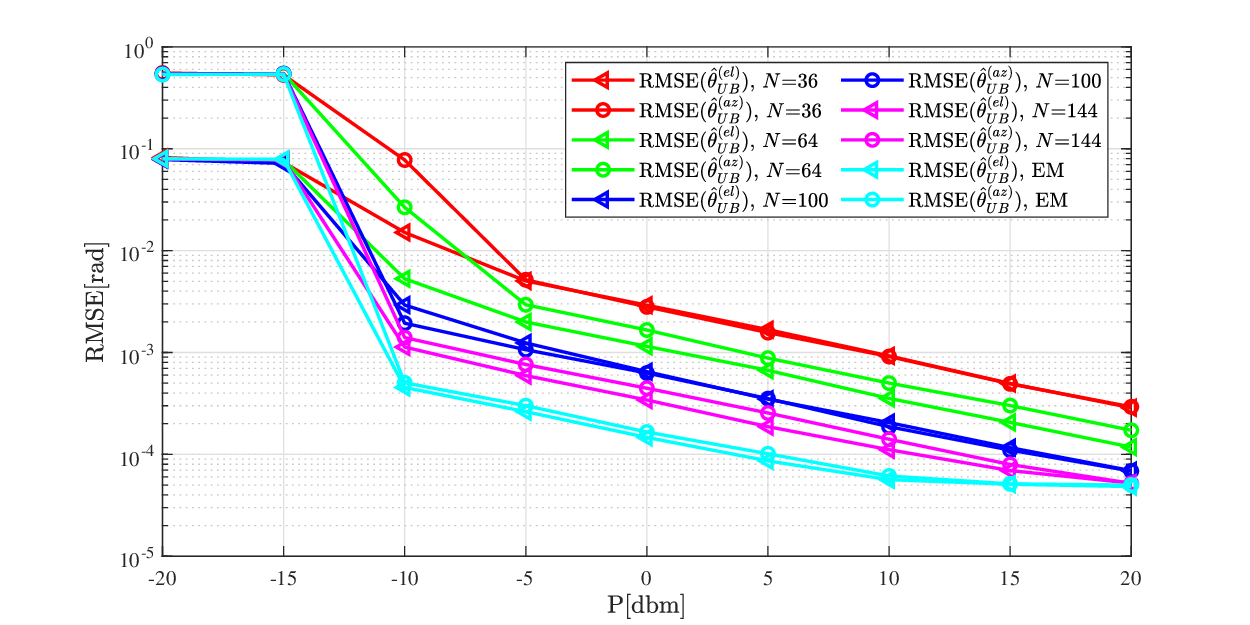}
		\caption{RMSE of ${{\bm{\theta }}_{{\text{UB}}}}$ versus number of FAS step sizes.}
		\label{fig4a}
	\end{figure}
	
	\begin{figure}[htbp]
		\centering
		\includegraphics[width=1.1\columnwidth]{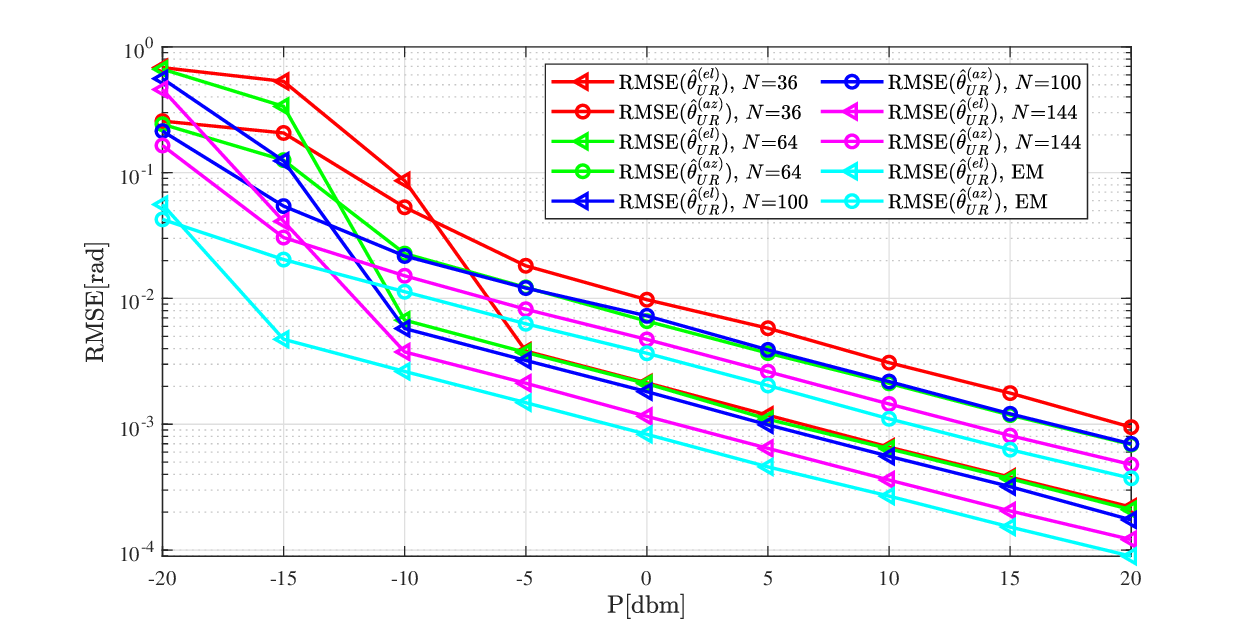}
		\caption{RMSE of ${{\bm{\theta }}_{{\text{UR}}}}$ versus number of FAS step sizes.}
		\label{fig4b}
	\end{figure}
	
	\begin{figure}[htbp]
		\centering
		\includegraphics[width=1.1\columnwidth]{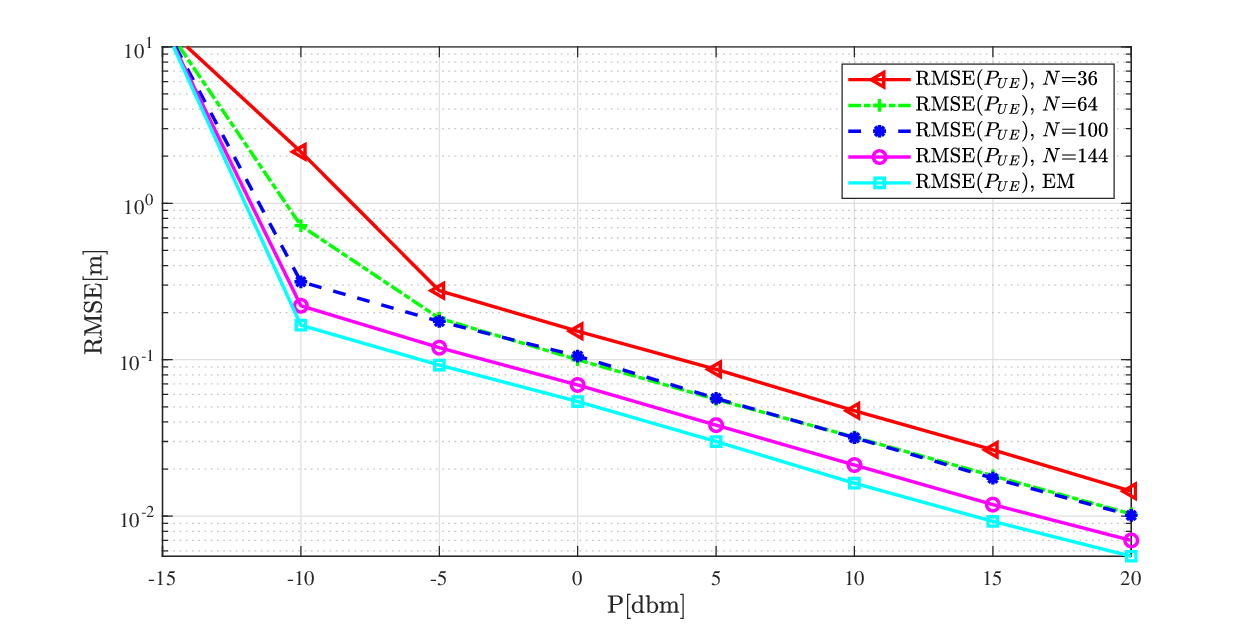}
		\caption{RMSE of ${{\mathbf{p}}_\textrm{U}}$ versus number of FAS step sizes.}
		\label{fig4c}
	\end{figure}
	
	\subsection{RMSE versus number of Scatterers}
	This subsection evaluates the algorithm's robustness to multipath propagation by introducing scatterers on the UE-RIS path ($L_{S1}$), ARIS-BS path ($L_{S2}$), and UE-BS path ($L_{S3}$). The scatterers are placed at fixed locations: ${\mathbf{p}}_{L_{S1}}^1 = {[-5.5, 28.6, 2]^T}$m, ${\mathbf{p}}_{L_{S1}}^2 = {[-2, 30, 3]^T}$m; ${\mathbf{p}}_{L_{S2}}^1 = {[-7, 8, 9.3]^T}$m, ${\mathbf{p}}_{L_{S2}}^2 = {[-6, 18.6, 2.7]^T}$m; and ${\mathbf{p}}_{L_{S3}}^1 = {[6.7, 28, 11]^T}$m, ${\mathbf{p}}_{L_{S3}}^2 = {[8, 5, 2]^T}$m.
	
	Fig.~\ref{fig5a} shows that the BS AoA estimation is highly robust. The performance curves are nearly identical regardless of the number or location of scatterers, indicating that the signal decoupling and MUSIC algorithm successfully isolate and estimate the strong LoS path.
	
	In stark contrast, Fig.~\ref{fig5b} shows that the ARIS AoA estimation is much more sensitive to multipath. The presence of scatterers on the UE-RIS link ($L_{S1}$) or the ARIS-BS link ($L_{S2}$) significantly degrades performance, creating an error floor at high SNR. This is because these multipath components directly interfere with the estimation of the cascaded channel parameters.
	
	As a result, the final localization accuracy, shown in Fig. \ref{fig5c}, is limited by the performance of the ARIS AoA estimation. The error floors in the presence of NLoS-link scatterers demonstrate this dependency.
	
	\begin{remark}
		The algorithm exhibits an asymmetric robustness to multipath.  This insight reinforces our earlier conclusion: the primary challenge and source of performance loss in this framework lies in accurately disentangling the parameters of the reflected path.
	\end{remark}
	
	\begin{figure}[htbp]
		\centering
		\includegraphics[width=1.1\columnwidth]{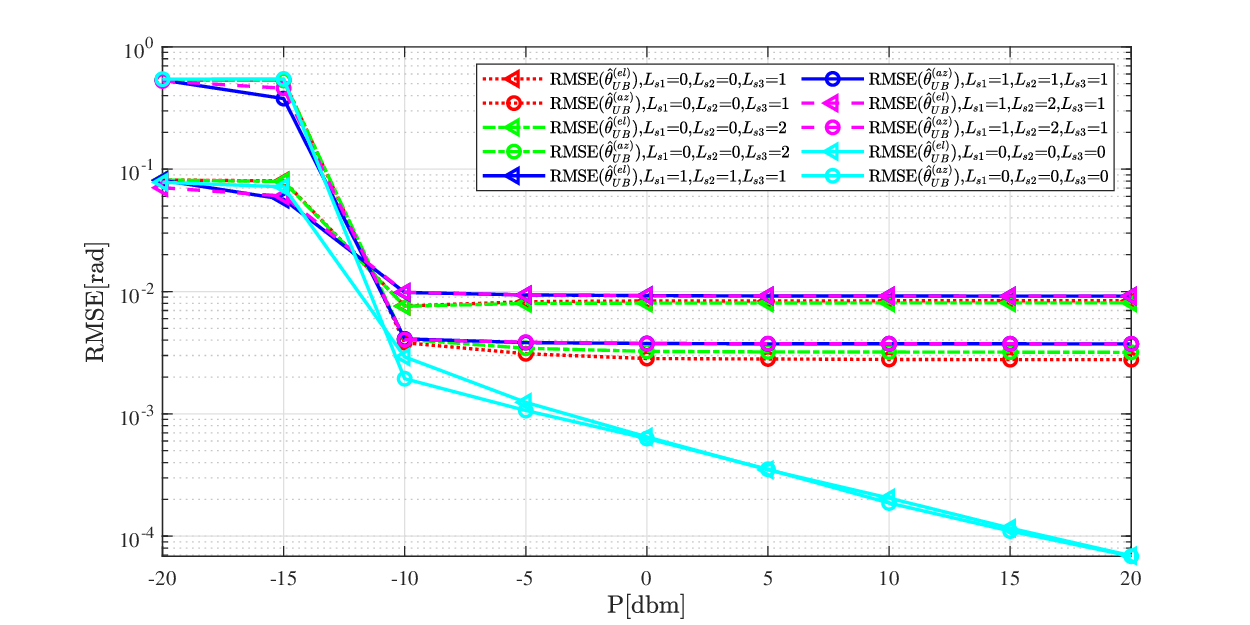}
		\caption{RMSE of ${{\bm{\theta }}_{{\text{UB}}}}$ versus number of Scatterers.}
		\label{fig5a}
	\end{figure}
	
	\begin{figure}[htbp]
		\centering
		\includegraphics[width=1.1\columnwidth]{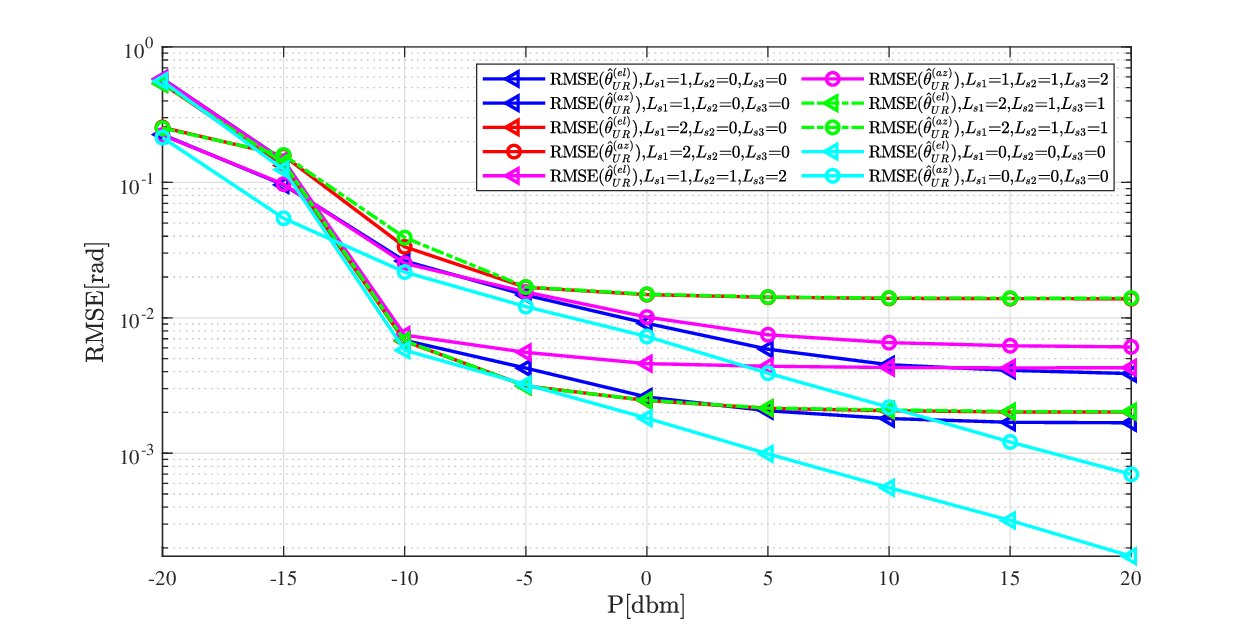}
		\caption{RMSE of ${{\bm{\theta }}_{{\text{UR}}}}$ versus number of Scatterers.}
		\label{fig5b}
	\end{figure}
	
	\begin{figure}[htbp]
		\centering
		\includegraphics[width=1.1\columnwidth]{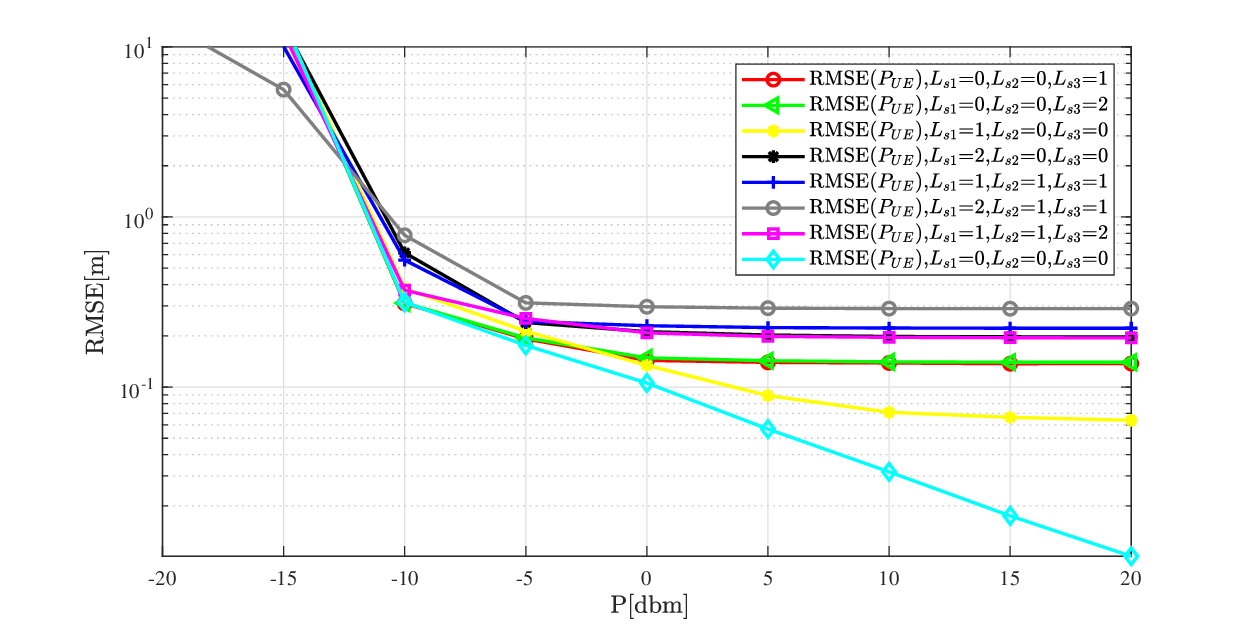}
		\caption{RMSE of ${{\mathbf{p}}_\textrm{U}}$ versus number of Scatterers.}
		\label{fig5c}
	\end{figure}
	
	\subsection{Performance Comparison with Passive RIS}
	This subsection compares the localization performance of the proposed ARIS-assisted model against a traditional passive RIS setup. Due to the inherent ``multiplicative fading" effect, a passive RIS is largely ineffective at low transmit powers. Therefore, this comparison focuses on the high-power regime where passive RIS can achieve a baseline localization performance, while the performance of ARIS at lower powers has already been established in Subsection VI-A. Since the LoS path is unaffected by the type of RIS, we omit the BS AoA results and focus on the RIS-dependent parameters.
	
	Fig.~\ref{fig6a} compares the AoA estimation RMSE at the RIS. The passive RIS only begins to function effectively for transmit powers above 30 {dbm}; below this threshold, its performance is poor and flat due to the low SNR of the reflected path. In contrast, the ARIS provides robust estimates across the power range. While it exhibits some performance saturation at very high powers, its estimation accuracy is consistently better than that of the passive RIS by approximately two orders of magnitude. {As the numbers of $N$ and $M_R$ increase, the estimation performances of FAS-ARIS improve consistently. The value of $\varepsilon$ that achieves optimal system performance also aligns closely with the value that maximizes the estimation performance of ${{\bm{\theta }}_{{\text{UR}}}}$.}
	
	The final UE positioning accuracy, shown in Fig. \ref{fig6b}, is determined by the joint precision of the AoA estimates. As the BS AoA estimation is identical for both cases, the positioning RMSE trend closely follows that of the RIS AoA. Consequently, the ARIS framework achieves a localization accuracy that is also about two orders of magnitude superior to the passive RIS counterpart.
	
	\begin{figure}[htbp]
		\centering
		\includegraphics[width=1.1\columnwidth]{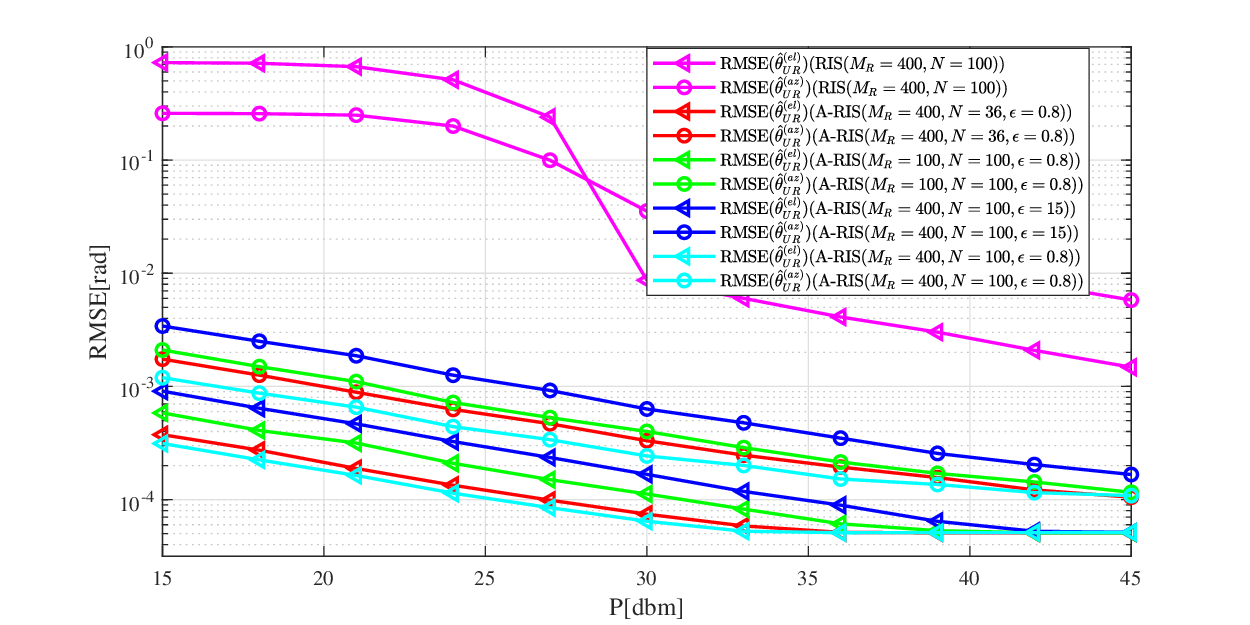}
		\caption{{RMSE of ${{\bm{\theta }}_{{\text{UR}}}}$ for Active vs. Passive RIS.}}
		\label{fig6a}
	\end{figure}
	
	\begin{figure}[htbp]
		\centering
		\includegraphics[width=1.1\columnwidth]{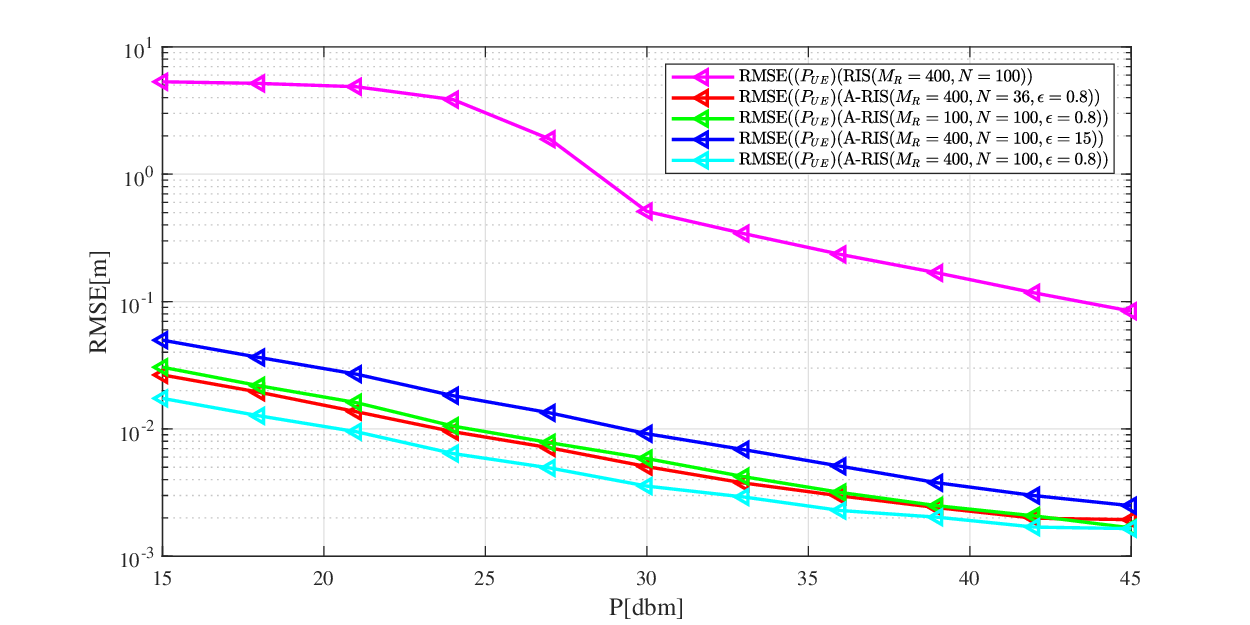}
		\caption{{RMSE of ${{\mathbf{p}}_\textrm{U}}$ for Active vs. Passive RIS.}}
		\label{fig6b}
	\end{figure}
	
	\section{Conclusion}\label{sec:conclude}
	In this paper, we proposed a wireless communication system for ARIS-assisted FAS to localize UE in multipath environments. Through the phase design of the ARIS and the frequency guide of the UE, we decouple the parameters of the coupling at the ARIS and separate the LOS and NLOS signals to estimate their channel parameters separately. For the LOS signal, we used the 2D MUSIC algorithm to get the AOA estimation at the BS. For NLOS signals, we used the same operation. First, combining the known position information of the BS and ARIS, the effect of the AOA at the BS of the NLOS channel is eliminated, and then the {cascaded} angle can be searched by constructing a MLE problem followed by a 2D search while the parameters are refined using the interior point method. Based on the {cascaded} angle definition, we recovered the AOA estimation of ARIS. Finally, the AOA of the BS and ARIS without scatterers, obtained from the above estimates, was used to estimate the UE's position using the LS method. Simulation results confirmed the effectiveness of the proposed method and showed that within a certain transmission power range, the RMSE of the channel parameter estimation reaches the corresponding CRBs.
	
	\appendix 
	\section{Appendix A}
	\subsection{FIM OF CHANNEL PARAMETERS}
	To obtain the partial derivatives of ${\bm{\gamma }}$ with respect to the channel parameters, differentiate (\ref{eq46}) with respect to each element of the channel parameter vector ${\bm{\gamma }}$ to obtain
	\begin{equation}\label{eq50}
		\frac{{\partial {{\bm{\mu }}_\textrm{n,t}}}}{{\partial \Re (\rho _\textrm{UB}^{})}} = {e^{ - j2\pi {{\mathbf{p}}_\textrm{B,n}}^T{\bm{k}}({\bm{\theta }}_{_\textrm{UB}}^{})d/\lambda }}{x_t},
	\end{equation}
	\begin{equation}\label{eq51}
		\frac{{\partial {{\bm{\mu }}_\textrm{n,t}}}}{{\partial \Im (\rho _\textrm{UB}^{})}} = j{e^{ - j2\pi {{\mathbf{p}}_\textrm{B,n}}^T{\bm{k}}({\bm{\theta }}_{_\textrm{UB}}^{})d/\lambda }}{x_t},
	\end{equation}
	\begin{align}\label{eq52}
		&\frac{\partial \bm{\mu}_\textrm{n,t}}{\partial \Re (\rho_\textrm{URB})}\nonumber\\
		&= \sum_{i=1}^{M_\textrm{R}} e^{-j\frac{2\pi}{\lambda} \left( \mathbf{p}_\textrm{B,n}^T \bm{k}(\bm{\theta}_\textrm{BR}) + \mathbf{p}_\textrm{R,i}^T \bm{k}(\bm{\theta}_\textrm{RB}) + \mathbf{p}_\textrm{R,1}^T \bm{k}(\bm{\theta}_\textrm{UR}^i) \right)}   \mathbf{w}_{t,i}   x_t,
	\end{align}
	
	\begin{align}\label{eq53}
		&\frac{\partial \bm{\mu}_\textrm{n,t}}{\partial \Im (\rho_\textrm{URB})}\nonumber\\
		&= j \sum_{i=1}^{M_\textrm{R}} e^{-j\frac{2\pi}{\lambda} \left( \mathbf{p}_\textrm{B,n}^T \bm{k}(\bm{\theta}_\textrm{BR}) + \mathbf{p}_\textrm{R,i}^T \bm{k}(\bm{\theta}_\textrm{RB}) + \mathbf{p}_\textrm{R,1}^T \bm{k}(\bm{\theta}_\textrm{UR}^i) \right)}  \mathbf{w}_{t,i}  x_t,
	\end{align}
	{
	\begin{align}\label{eq54}
		\frac{\partial \bm{\mu}_\textrm{n,t}}{\partial \bm{\theta}_\textrm{UB}^\textrm{el}}
		= -j\frac{2\pi}{\lambda}   \mathbf{p}_\textrm{B,n}^T   \rho_\textrm{UB}   e^{-j\frac{2\pi}{\lambda} \mathbf{p}_\textrm{B,n}^T \bm{k}(\bm{\theta}_\textrm{UB})}   x_t   \frac{\partial \bm{k}(\bm{\theta}_\textrm{UB})}{\partial \bm{\theta}_\textrm{UB}^\textrm{el}},
	\end{align}
	\begin{align}\label{eq55}
		\frac{\partial \bm{\mu}_\textrm{n,t}}{\partial \bm{\theta}_\textrm{UB}^\textrm{az}}
		= -j\frac{2\pi}{\lambda}   \mathbf{p}_\textrm{B,n}^T   \rho_\textrm{UB}   e^{-j\frac{2\pi}{\lambda} \mathbf{p}_\textrm{B,n}^T \bm{k}(\bm{\theta}_\textrm{UB})}   x_t   \frac{\partial \bm{k}(\bm{\theta}_\textrm{UB})}{\partial \bm{\theta}_\textrm{UB}^\textrm{az}},
	\end{align}
	\begin{align}\label{eq56}
		&\frac{\partial \bm{\mu}_\textrm{n,t}}{\partial \bm{\theta}_\textrm{UR}^\textrm{el}} = -j\frac{2\pi \mathbf{p}_\textrm{R,i}^T \rho_\textrm{URB}}{\lambda} \sum_{i=1}^{M_\textrm{R}} e^{-j\frac{2\pi}{\lambda} \left( \mathbf{p}_\textrm{B,n}^T \bm{k}(\bm{\theta}_\textrm{BR}) + \mathbf{p}_\textrm{R,i}^T \bm{k}(\bm{\theta}_\textrm{RB}) + \mathbf{p}_\textrm{R,1}^T \bm{k}(\bm{\theta}_\textrm{UR}) \right)} \nonumber\\
		&\qquad \qquad\cdot \mathbf{w}_{t,i} \, x_t \, \frac{\partial \bm{k}(\bm{\theta}_\textrm{UR})}{\partial \bm{\theta}_\textrm{UR}^\textrm{el}},
	\end{align}
	\begin{align}\label{eq57}
		&\frac{\partial \bm{\mu}_\textrm{n,t}}{\partial \bm{\theta}_\textrm{UR}^\textrm{az}} = -j\frac{2\pi \mathbf{p}_\textrm{R,i}^T \rho_\textrm{URB}}{\lambda} \sum_{i=1}^{M_\textrm{R}} e^{-j\frac{2\pi}{\lambda} \left( \mathbf{p}_\textrm{B,n}^T \bm{k}(\bm{\theta}_\textrm{BR}) + \mathbf{p}_\textrm{R,i}^T \bm{k}(\bm{\theta}_\textrm{RB}) + \mathbf{p}_\textrm{R,1}^T \bm{k}(\bm{\theta}_\textrm{UR}) \right)} \nonumber\\
		& \qquad\qquad\cdot \mathbf{w}_{t,i} \, x_t \, \frac{\partial \bm{k}(\bm{\theta}_\textrm{UR})}{\partial \bm{\theta}_\textrm{UR}^\textrm{az}},
	\end{align}
}
	
	Substituting these partial derivatives into (\ref{eq45}) yields the CRB for channel parameter estimation in equation (\ref{eq47}).
	
	\section{Appendix B}
	\subsection{DERIVATION OF JACOBIAN MATRIX}
	According to the definition of the Jacobian matrix, the following derivatives can be obtained as
	\begin{align}\label{eq58}
		\mathbf{J}(1,1) &= \frac{\partial \Re (\rho_\textrm{UB})}{\partial \Re (\rho_\textrm{UB})} = 1, \\
		\mathbf{J}(2,2) &= \frac{\partial \Im (\rho_\textrm{UB})}{\partial \Im (\rho_\textrm{UB})} = 1 , \\
		\mathbf{J}(3,3) &= \frac{\partial \Re (\rho_\textrm{URB})}{\partial \Re (\rho_\textrm{URB})} = 1,  \\
		\mathbf{J}(4,4) &= \frac{\partial \Im (\rho_\textrm{URB})}{\partial \Im (\rho_\textrm{URB})} = 1
	\end{align}
	\begin{equation}\label{eq59}
		{\mathbf{J}}(5,5) = \frac{{\partial ({\mathbf{\theta }}_\textrm{UB}^\textrm{el})}}{{\partial {{\mathbf{p}}_\textrm{U}}(1)}} = \frac{{({x_\textrm{U}} - {x_\textrm{B}})({z_\textrm{U}} - {z_\textrm{B}})}}{{\sqrt {{{({x_\textrm{U}} - {x_\textrm{B}})}^2} + {{({y_\textrm{U}} - {y_\textrm{B}})}^2}} {{\left\| {{{\mathbf{p}}_\textrm{U}} - {{\mathbf{p}}_\textrm{B}}} \right\|}^2}}}
	\end{equation}
	{
	\begin{equation}\label{eq60}
		{\mathbf{J}}(5,6) = \frac{{\partial ({\mathbf{\theta }}_\textrm{UB}^\textrm{el})}}{{\partial {{\mathbf{p}}_\textrm{U}}(2)}} = \frac{{({y_\textrm{U}} - {y_\textrm{B}})({z_\textrm{U}} - {z_\textrm{B}})}}{{\sqrt {{{({x_\textrm{U}} - {x_\textrm{B}})}^2} + {{({y_\textrm{U}} - {y_\textrm{B}})}^2}} {{\left\| {{{\mathbf{p}}_\textrm{U}} - {{\mathbf{p}}_\textrm{B}}} \right\|}^2}}}
	\end{equation}}
	\begin{equation}\label{eq61}
		{\mathbf{J}}(5,7) = \frac{{\partial ({\mathbf{\theta }}_\textrm{UB}^\textrm{el})}}{{\partial {{\mathbf{p}}_\textrm{U}}(3)}} =  - \frac{{\sqrt {{{({x_\textrm{U}} - {x_\textrm{B}})}^2} + {{({y_\textrm{U}} - {y_\textrm{B}})}^2}} }}{{{{\left\| {{{\mathbf{p}}_\textrm{U}} - {{\mathbf{p}}_\textrm{B}}} \right\|}^2}}}
	\end{equation}
	\begin{equation}\label{eq62}
		{\mathbf{J}}(6,5) = \frac{{\partial ({\mathbf{\theta }}_\textrm{UB}^\textrm{az})}}{{\partial {{\mathbf{p}}_\textrm{U}}(1)}} = \frac{{ - ({y_\textrm{U}} - {y_\textrm{B}})}}{{{{({x_\textrm{U}} - {x_\textrm{B}})}^2} + {{({y_\textrm{U}} - {y_\textrm{B}})}^2}}}
	\end{equation}
	\begin{equation}\label{eq63}
		{\mathbf{J}}(6,6) = \frac{{\partial ({\mathbf{\theta }}_\textrm{UB}^\textrm{az})}}{{\partial {{\mathbf{p}}_\textrm{U}}(2)}} = \frac{{{x_\textrm{U}} - {x_\textrm{B}}}}{{{{({x_\textrm{U}} - {x_\textrm{B}})}^2} + {{({y_\textrm{U}} - {y_\textrm{B}})}^2}}}
	\end{equation}
	\begin{equation}\label{eq64}
		{\mathbf{J}}(7,5) = \frac{{\partial ({\mathbf{\theta }}_\textrm{UR}^\textrm{el})}}{{\partial {{\mathbf{p}}_\textrm{U}}(1)}} = \frac{{({x_\textrm{U}} - {x_\textrm{R}})({z_\textrm{U}} - {z_\textrm{R}})}}{{\sqrt {{{({x_\textrm{U}} - {x_\textrm{R}})}^2} + {{({y_\textrm{U}} - {y_\textrm{R}})}^2}} {{\left\| {{{\mathbf{p}}_\textrm{U}} - {{\mathbf{p}}_\textrm{R}}} \right\|}^2}}}
	\end{equation}
	\begin{equation}\label{eq65}
		{\mathbf{J}}(7,6) = \frac{{\partial ({\mathbf{\theta }}_\textrm{UR}^\textrm{el})}}{{\partial {{\mathbf{p}}_\textrm{U}}(2)}} = \frac{{({y_\textrm{U}} - {y_\textrm{R}})({z_\textrm{U}} - {z_\textrm{R}})}}{{\sqrt {{{({x_\textrm{U}} - {x_\textrm{R}})}^2} + {{({y_\textrm{U}} - {y_\textrm{R}})}^2}} {{\left\| {{{\mathbf{p}}_\textrm{U}} - {{\mathbf{p}}_\textrm{R}}} \right\|}^2}}}
	\end{equation}
	\begin{equation}\label{eq66}
		{\mathbf{J}}(7,7) = \frac{{\partial ({\mathbf{\theta }}_\textrm{UR}^\textrm{el})}}{{\partial {{\mathbf{p}}_\textrm{U}}(3)}} =  - \frac{{\sqrt {{{({x_\textrm{U}} - {x_\textrm{R}})}^2} + {{({y_\textrm{U}} - {y_\textrm{R}})}^2}} }}{{{{\left\| {{{\mathbf{p}}_\textrm{U}} - {{\mathbf{p}}_\textrm{R}}} \right\|}^2}}}
	\end{equation}
	\begin{equation}\label{eq67}
		{\mathbf{J}}(8,5) = \frac{{\partial ({\mathbf{\theta }}_\textrm{UR}^\textrm{az})}}{{\partial {{\mathbf{p}}_\textrm{U}}(1)}} = \frac{{ - ({y_\textrm{U}} - {y_\textrm{R}})}}{{{{({x_\textrm{U}} - {x_\textrm{R}})}^2} + {{({y_\textrm{U}} - {y_\textrm{R}})}^2}}}
	\end{equation}
	\begin{equation}\label{eq68}
		{\mathbf{J}}(8,6) = \frac{{\partial ({\mathbf{\theta }}_\textrm{UR}^\textrm{az})}}{{\partial {{\mathbf{p}}_\textrm{U}}(2)}} = \frac{{{x_\textrm{U}} - {x_\textrm{R}}}}{{{{({x_\textrm{U}} - {x_\textrm{R}})}^2} + {{({y_\textrm{U}} - {y_\textrm{R}})}^2}}}
	\end{equation}
	where ${{\mathbf{p}}_\textrm{B}} = {[{x_\textrm{B}},{y_\textrm{B}},{z_\textrm{B}}]^{{T}}},{{\mathbf{p}}_\textrm{R}} = {[{x_\textrm{R}},{y_\textrm{R}},{z_\textrm{R}}]^{{T}}}$, the remaining elements of the Jacobi matrix are zero.

	
	%

	%
	%
	%

	\ifCLASSOPTIONcaptionsoff
	\newpage
	\fi
	\bibliographystyle{IEEEtran}

\end{document}